\documentclass[12pt,preprint]{emulateapj}

\newcommand{\CIV}{\hbox{{\rm C}\kern 0.1em{\sc iv}}}
\newcommand{\HeII}{\hbox{{\rm He}\kern 0.1em{\sc ii}\kern 0.1em{$\lambda1640$}}}
\newcommand{\OIII}{\hbox{{\rm O}\kern 0.1em{\sc iii}}]}
\begin{document}
\title{Newly Discovered Bright $z\sim9$-10 Galaxies and Improved
  Constraints on Their Prevalence Using the Full CANDELS Area}
\author{Bouwens, R.J.\altaffilmark{1}, Stefanon, M.\altaffilmark{1},
  Oesch, P.A.\altaffilmark{2,3}, Illingworth, G.D.\altaffilmark{4}, Nanayakkara, T.\altaffilmark{1}, Roberts-Borsani, G.\altaffilmark{5}, Labb{\'e}, I.\altaffilmark{6},  Smit, R.\altaffilmark{7}}
\altaffiltext{1}{Leiden Observatory, Leiden University,
  NL-2300 RA Leiden, Netherlands}
\altaffiltext{2}{Observatoire de Gen{\`e}ve, 1290 Versoix, Switzerland}
\altaffiltext{3}{International Associate, Cosmic Dawn Center (DAWN) at the Niels Bohr Institute, University of Copenhagen and DTU-Space, Technical University of Denmark}
\altaffiltext{4}{UCO/Lick Observatory, University of California, Santa
  Cruz, CA 95064}
\altaffiltext{5}{Astrophysics Group, Department of Physics and Astronomy, University College London, Gower Street, London, WC1E 6BT}
\altaffiltext{6}{Centre for Astrophysics \& Supercomputing, Swinburne University of Technology, PO Box 218, Hawthorn, VIC 3112, Australia}
\altaffiltext{7}{Cavendish Laboratory, University of Cambridge, 19 JJ Thomson Avenue, Cambridge CB3 0HE, UK}
\begin{abstract}
We report the results of an expanded search for $z\sim9$-10 candidates
over the $\sim$883 arcmin$^2$ CANDELS+ERS fields.  This study adds 147
arcmin$^2$ to the search area we consider over the CANDELS COSMOS,
UDS, and EGS fields, while expanding our selection to include sources
with bluer $J_{125}-H_{160}$ colors than our previous
$J_{125}-H_{160}>0.5$ mag selection.  In searching for new $z\sim9$-10
candidates, we make full use of all available HST, Spitzer/IRAC, and
ground-based imaging data.  As a result of our expanded search and use
of broader color criteria, 3 new candidate $z\sim9$-10 galaxies are
identified.  We also find again the $z=8.683$ source previously
confirmed by Zitrin et al.\ (2015).  This brings our sample of
probable $z\sim9$-11 galaxy candidates over the CANDELS+ERS fields to
19 sources in total, equivalent to 1 candidate per 47 arcmin$^2$ (1
per 10 WFC3/IR fields).  To be comprehensive, we also discuss 28
mostly lower likelihood $z\sim9$-10 candidates, including some sources
that seem to be reliably at $z>8$ using the HST+IRAC data alone, but
which the ground-based data show are much more likely at $z<4$.  One
case example is a bright $z\sim9.4$ candidate COS910-8 which seems
instead to be at $z\sim2$.  Based on this expanded sample, we obtain a
more robust LF at $z\sim9$ and improved constraints on the volume
density of bright $z\sim9$ and $z\sim10$ galaxies.  Our improved
$z\sim9$-10 results again reinforce previous findings for strong
evolution in the $UV$ LF at $z>8$, with a factor of $\sim$10 evolution
seen in the luminosity density from $z\sim10$ to $z\sim8$.
\end{abstract}

\section{Introduction}

Over the last few years, the search for galaxies in the early universe
has revealed sources out to redshifts as high as $z\sim11$ (Coe et
al.\ 2013; Oesch et al.\ 2016), corresponding to 400 million years
after the Big Bang.  Simultaneous with these activities, tens of
galaxies have been identified some 50-150 Myr later than this, at
$z\sim9$-10 (Bouwens et al.\ 2011; Zheng et al.\ 2012; Coe et
al.\ 2013; Ellis et al.\ 2013; McLure et al.\ 2014; Oesch et
al.\ 2013, 2014; Zitrin et al.\ 2014; Bouwens et al.\ 2015, 2016;
McLeod et al.\ 2015, 2016; Ishigaki et al.\ 2018; Oesch et al.\ 2018).

In the search for distant galaxies, one surprise was the discovery of
very bright ($M_{UV,AB}\lesssim -22$) galaxies at $z\sim9$-11 (Oesch
et al.\ 2014).  Subsequent work within the HST Cosmic Assembly
Near-Infrared Deep Extragalactic Legacy Survey (CANDELS) and Early
Release Science (ERS) programs (Grogin et al.\ 2011; Koekemoer et
al.\ 2011; Windhorst et al.\ 2011) and also pure parallel HST programs
like BoRG/HIPPIES (Yan et al.\ 2011; Trenti et al.\ 2011) have added
to the number of bright ($M_{UV,AB}\sim-21$) $z\sim9$-10 candidates
known (Bouwens et al.\ 2016; Calvi et al.\ 2016; Bernard et al.\ 2016;
Livermore et al.\ 2018; Morishita et al.\ 2018).  Of the known bright
$z>9$ galaxies, the most extreme example has been the $z=11.1\pm0.1$
galaxy GN-z11 (Oesch et al.\ 2016), which owing to its exceptional
brightness ($M_{UV,AB}\lesssim-22$) and high redshift must have
required an especially rare, overdense region of the universe to form
(Mutch et al.\ 2016; Waters et al.\ 2016).

Identifications of such bright galaxies have been useful not only
because of the amenability of the sources for spectroscopic follow-up
work and redshift determinations (Zitrin et al.\ 2015; Oesch et
al.\ 2016; Hashimoto et al.\ 2018), but also because of the utility of
such sources to further characterization, i.e., allowing for
properties like their dust content (Watson et al.\ 2015), dynamical
properies (Smit et al.\ 2018), $UV$-continuum slopes (Wilkins et
al.\ 2016), stellar masses (Lam et al.\ 2019), and physical sizes
(Holwerda et al.\ 2015) to be examined in detail.

Despite significant work done to the present in searching for
$z\sim9$-10 galaxies, we can still make progress in expanding current
$z\sim9$-10 samples using existing data sets.  For example, in the
Bouwens et al.\ (2016) search for $z\sim9$-10 candidate galaxies over
the CANDELS fields, consideration was only given to those WFC3/IR
regions with deep optical observations from the CANDELS program, i.e.,
roughly $\sim$85\% of the CANDELS area.  In addition, Bouwens et
al.\ (2016) focused on galaxies with the reddest $J_{125}-H_{160}$
colors in obtaining follow-up observations with HST (Bouwens 2014: GO
13752) both to search for the highest redshift sources ($z\gtrsim8.9$)
and to maximize the efficiency of the follow-up observations.  By
adopting this conservative approach, however, Bouwens et al.\ (2016)
potentially missed some $z\lesssim8.9$ galaxies with somewhat bluer
$J_{125}-H_{160}$ colors.

Here, we expand the search for $z\sim9$-10 galaxies to include the
full $\sim$883 arcmin$^2$ area within CANDELS+ERS.  Our new search
includes a 147 arcmin$^2$ area with deep WFC3/IR observations not
utilized in previous work.  We have expanded the area we consider
within CANDELS, mostly by leveraging ground-based observations where
deep ACS/optical data are not available.  Our new search results also
benefit from our considering sources with a broader set of
$J_{125}-H_{160}$ colors than we had previously considered and
inclusion of some additional HST follow-up observations taken in cycle
23 (GO 14459: Bouwens 2015).  As part of our expanded search, we also
pursue the selection of $z\sim9$-10 sources using the HST and
Spitzer/IRAC 3.6$\mu$m+4.5$\mu$m data alone, in case confusion in the
ground-based data resulted in our missing some sources in our earlier
study (Bouwens et al.\ 2016).

\begin{figure*}
\epsscale{1.15}
\plotone{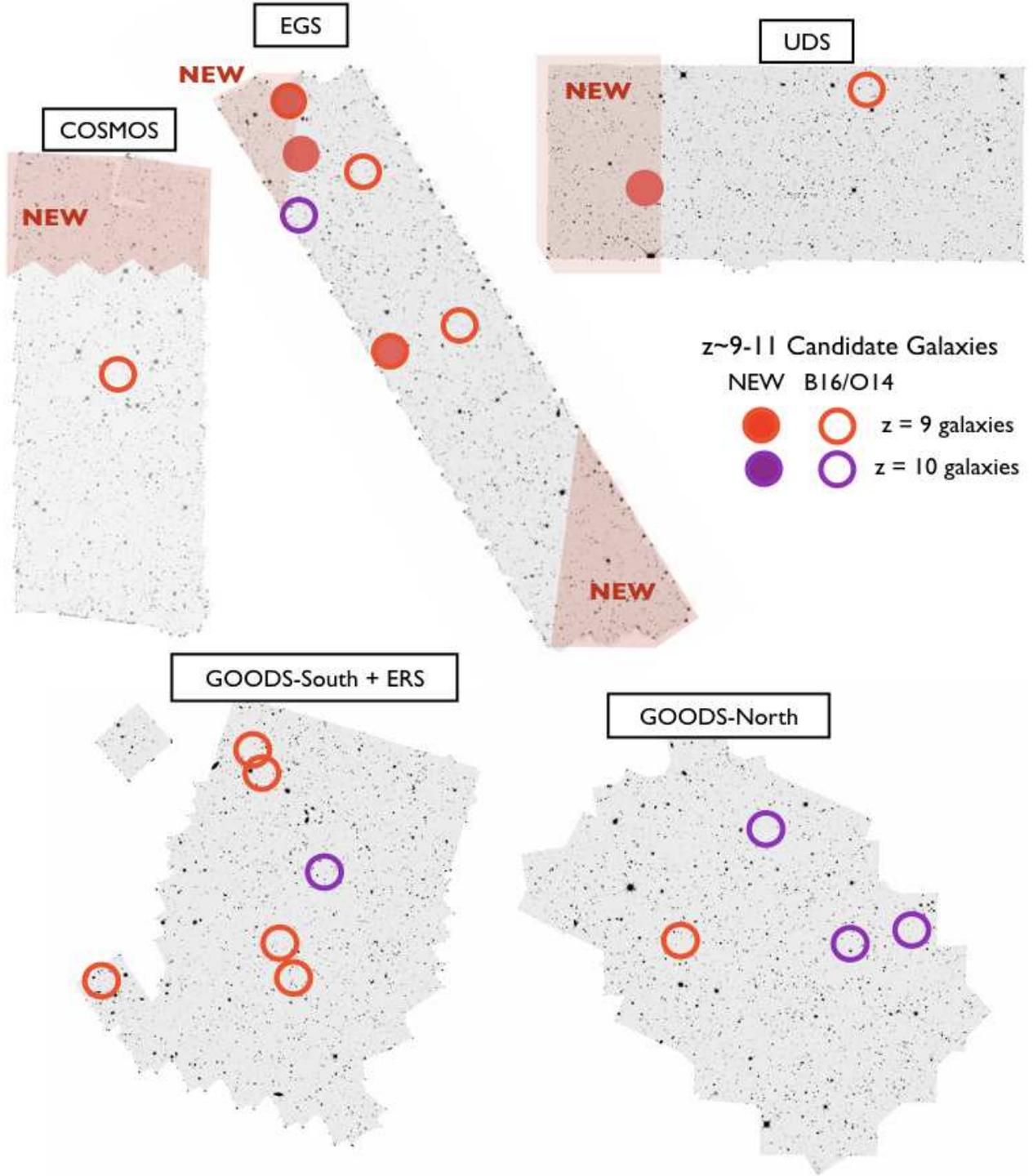}
\caption{Observational footprints showing the layout of the sensitive
  $J_{125}$ and $H_{160}$ WFC3/IR observations over the five CANDELS
  fields.  The regions enclosed by the red lines indicate the new
  WFC3/IR areas within CANDELS where searches for $z\sim9$-10 galaxies
  are performed in this study.  In our earlier study (Bouwens et
  al.\ 2016), optical ACS observations were not available to us when
  we performed our earlier study, and so these regions were not
  considered.  Here ground-based data are utilized when deep ACS
  $V_{606}$ and $I_{814}$ data from CANDELS are not available.  These
  regions correspond to 53.4 arcmin$^2$, 48.7 arcmin$^2$, and 45.3
  arcmin$^2$ over the CANDELS EGS, COSMOS, and UDS fields,
  respectively, for a total area of 147 arcmin$^2$. The solid red
  circles show the position of new ($P(z>8)$$>$0.5) $z\sim9$-10
  candidate galaxies identified by our selection criteria, while the
  open red and violet circles show the position of similar sources
  identified in our earlier studies (Oesch et al.\ 2014; Bouwens et
  al.\ 2015, 2016).\label{fig:layout}}
\end{figure*}

The plan for this paper is as follows.  In \S2, we provide a brief
description of the observational data we utilize and our procedures
for performing photometry.  In \S3, we describe our selection
procedure and results, while taking advantage of some cycle 23
observations to refine our constraints on the redshift of two $z\sim9$
candidates we had identified.  In \S4, we take advantage of the new
results to obtain improved estimates of the bright ends of the
$z\sim9$ and $z\sim10$ LFs.  We also attempt to quantify variations in
the volume densities of bright $z\sim9$ galaxies across the CANDELS
fields.  Finally, \S5 summarizes the results of this paper.  With
Appendices A and B, we consider results from our HST follow-up program
in cycle 23, while relaxing further our selection criteria for
identifying $z\sim9$-10 galaxies in the interests of constructing a
more complete sample of such sources.

For convenience, we frequently write the HST F606W, F814W, F098M,
F125W, F140W, and F160W filters as $V_{606}$, $I_{814}$, $Y_{098}$,
$J_{125}$, $JH_{140}$, and $H_{160}$, respectively, throughout this
work.  Motivated by recent Planck results and for consistency with
previous observational work (Planck Collaboration 2018), all results
here are presented in terms of the standard concordance cosmology,
with $\Omega_{m} = 0.3$, $\Omega_{\Lambda} = 0.7$, and $H_0 = 70$
km/s/Mpc.  We adopt the AB magnitude system (Oke \& Gunn 1983) throughout.

\begin{figure}
\epsscale{1.15}
\plotone{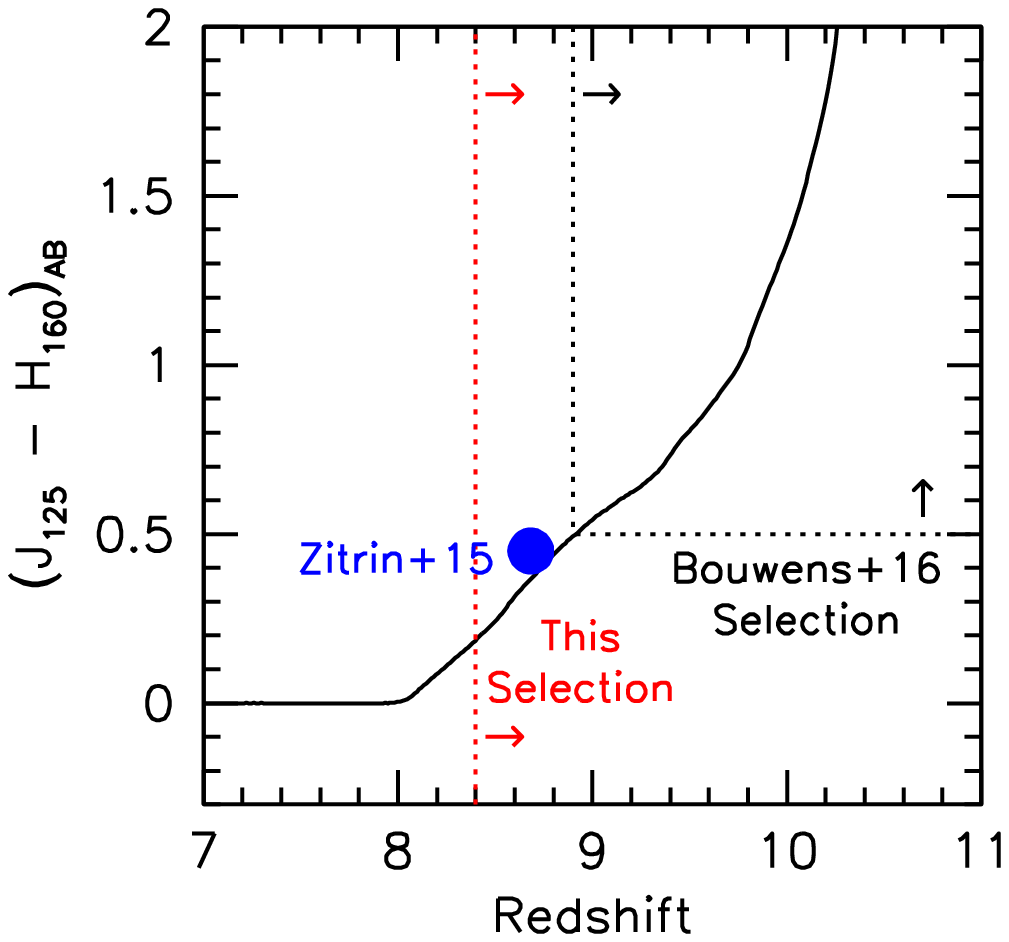}
\caption{Expected $J_{125}-H_{160}$ colors for a star-forming galaxy
  with $UV$-continuum slope $-2$ vs. redshift.  The red dotted lines
  and arrows indicate our inclusion of all sources with a photometric
  redshift in excess of 8.4.  The new selection criteria are shown
  relative to the $J_{125}-H_{160}>0.5$ color criterion used in
  Bouwens et al.\ (2016: \textit{black dotted lines and arrow}) to
  identify source for follow-up observations.  The redshift and
  measured $J_{125}-H_{160}$ color of the Zitrin et al.\ (2015) source
  is shown with the solid blue circle for context.  Bouwens et
  al.\ (2016) focused on sources with $J_{125}-H_{160}$ colors $>$0.5
  to maximize the efficiency of their follow-up observations, but this
  resulted in their being more incomplete regarding their
  identification of sources in the redshift range $z\sim8.4$-8.9.
  Here we make use of more inclusive selection criteria to identify a
  larger number of star-forming galaxies at
  $z\gtrsim8.4$.\label{fig:selc}}
\end{figure}

\section{Observational Data and Photometry}

Here we make use of a $\sim$883 arcmin$^2$ area with the five
CANDELS+ERS fields to search for bright candidate $z\sim9$-10
galaxies.  Our total search area includes both previously searched
regions of CANDELS (736 arcmin$^2$: \S2.2) and new search area
($\sim$147 arcmin$^2$: \S2.1).

\subsection{New Search Area Within CANDELS}

Here we make use of an additional $\sim$147 arcmin$^2$ search area
within CANDELS not considered in our earlier studies (Bouwens et
al.\ 2015, 2016).  Like the previously searched regions, the new
region also features deep ($\sim$26.5 mag, $5\sigma$) $J_{125}$ +
$H_{160}$ observations.  Observations of this depth are of significant
utility for finding $z\sim9$-10 galaxy candidates, given the
increasing prevalence of such candidates at $\gtrsim$25 mag and
especially $\gtrsim$26 mag.

The primary reason we did not consider this area in Bouwens et
al.\ (2016) was because of the lack of especially deep ACS/optical
observations over much of the area.  The extreme outermost regions of
the EGS mosaic were also not considered in our earlier search due to
our lacking reductions of the HST data in those areas when devising
our HST program to follow up specific $z\sim9$-10 candidates (GO 13792:
Bouwens 2014).

As with the case of the new EGS area, the regions considered here are
located towards the edges of the CANDELS UDS and COSMOS fields (due to
the roll angle constraints in scheduling HST observations and
arranging for the ACS/optical parallels to land on the WFC3/IR
observations).  The new regions of the CANDELS UDS, COSMOS, and EGS
mosaics are indicated using the light red shaded regions in
Figure~\ref{fig:layout}.  The new areas probed in each field subtend
45.3, 48.7, 53.4 arcmin$^2$, respectively, or 147.4 arcmin$^2$ in
total.

\begin{figure*}
\epsscale{1.15}
\plotone{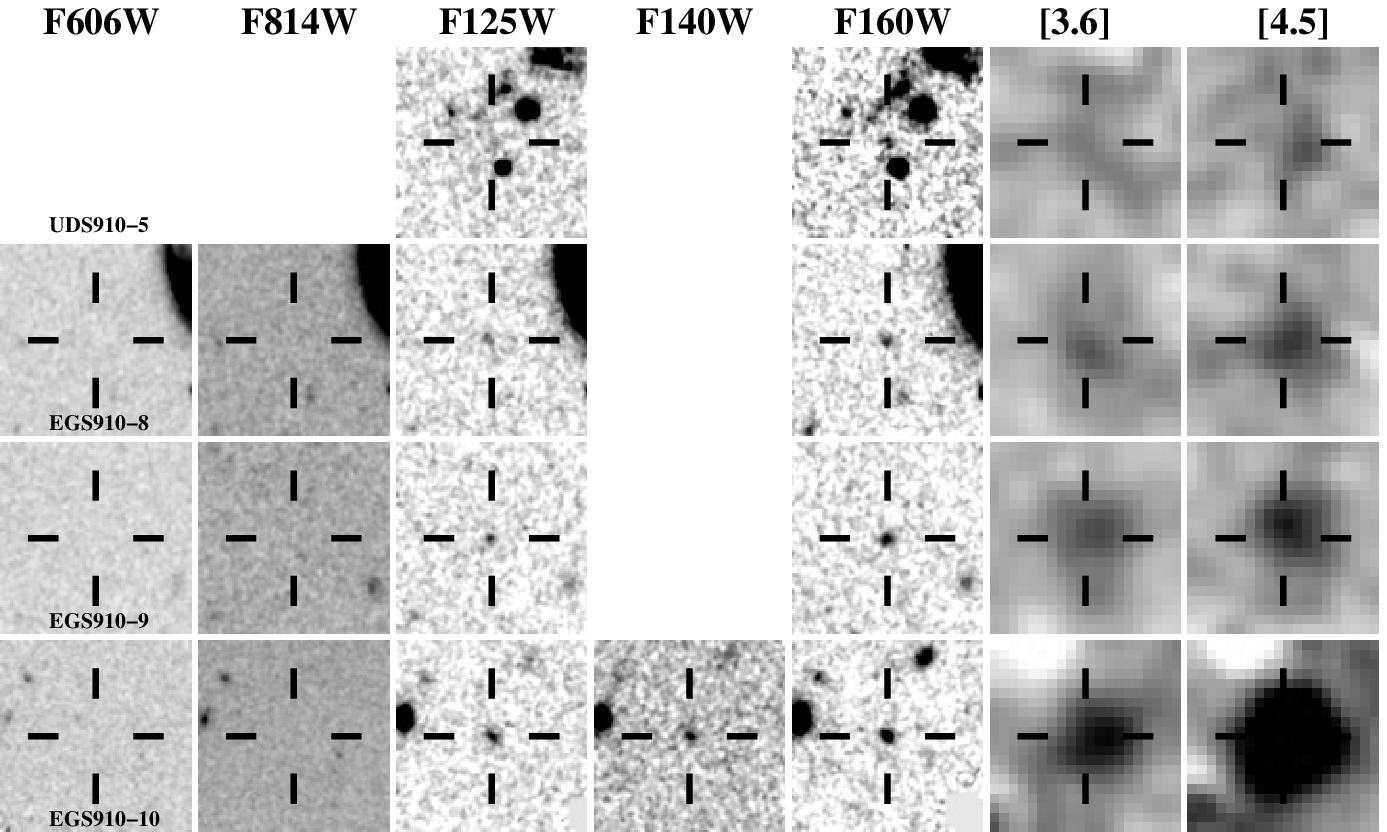}
\caption{Postage stamp images of the $P(z>8)>0.5$ $z\sim9$-10 galaxy
  candidates we have identified.  HST and Spitzer/IRAC images of the
  candidates are presented (where available) from left to right in the
  $V_{606}$, $I_{814}$, $J_{125}$, $JH_{140}$, $H_{160}$, [3.6], and
  [4.5] bands.  The presented postage stamps for the two Spitzer/IRAC
  bands is after subtraction of flux from the neighbors.  Each of the
  presented sources is securely detected in the $H_{160}$ band (and
  also the $JH_{140}$ band when available), but not in the optical ACS
  $V_{606}$ or $I_{814}$ bands.\label{fig:stamps}}
\end{figure*}

In constructing catalogs over these new WFC3/IR areas, we make use of
the reductions from the 3D-HST team (Skelton et al.\ 2014) which are
drizzled onto a 0.06$''$ grid.  Despite the lack of ACS coverage over
the northern CANDELS COSMOS region from CANDELS, such coverage is
available in the $I_{814}$ band thanks to the original COSMOS program
(Scoville et al.\ 2007).  Deep ACS optical $V_{606}$ and
$I_{814}$-band coverage is available from the original CANDELS program
over most of the new CANDELS EGS regions we search.  The v1.2
reductions of the COSMOS ACS data (Koekemoer et al.\ 2007) and CANDELS
EGS ACS data (Koekemoer et al.\ 2011) were retrieved from MAST and
registered against the WFC3/IR observations.

In addition, we use ground-based observations over the new CANDELS
regions to improve our constraints on the photometric redshifts of
sources.  Over the UDS, COSMOS, and EGS fields, use was made of the
Cirasuolo et al.\ (2010) reductions of deep optical Subaru Suprime-Cam
UDS/SXDS observations (Furusawa et al. 2008), the version 7 reductions
of the CFHTLS survey
observations\footnote{http://www.cfht.hawaii.edu/Science/CFHTLS}, and
the very deep reductions (Capak et al.\ 2007) of the Subaru
observations in the $B$, $g$, $V$, $r$, $i$, and $z$ bands.  At
near-IR wavelengths, use of the especially sensitive DR3 observations
over COSMOS with UltraVISTA in the $YJHK_s$ bands (McCracken et
al.\ 2012), the sensitive version 7 UKIRT/WFCAM $JHK_s$ observations
over UDS, and CFHT/WIRCam $K_s$ observations over EGS field (McCracken
et al.\ 2010; Bielby et al.\ 2012).  The depths of the optical data
reach to $\sim$27-28 mag ($5\sigma$) and in the near-IR, these data
reach to $\sim$25-26 mag ($5\sigma$).  

Finally, for Spitzer/IRAC, use is made of the Spitzer/IRAC S-CANDELS
and SEDS observations (Ashby et al.\ 2013, 2015), as well as any other
Spitzer/IRAC observations available over the CANDELS regions.
Reductions of these data were performed as detailed in Labb{\'e} et
al.\ (2015).

\subsection{Previously Searched CANDELS Areas}

During the process of considering a larger search area within CANDELS,
we also took the opportunity to conduct a second search for
$z\sim9$-10 candidates in regions already considered in Bouwens et
al.\ (2016).  We utilize the same reductions of the WFC3/IR and ACS
observations as were presented in Bouwens et al.\ (2016).  Those data
sets reach to $\sim$26.5 mag in the WFC3/IR bands ($5\sigma$) and
$\sim$27.0 mag in the optical bands.  As in our previous work, use of
the available ground-based + Spitzer/IRAC observations is made to
obtain the best constraints on any candidate $z\sim9$-10 galaxies that
are identified.  The properties of these data sets are as described in
the previous subsection (see also Bouwens et al.\ 2016).

\begin{figure*}
\epsscale{0.9}
\plotone{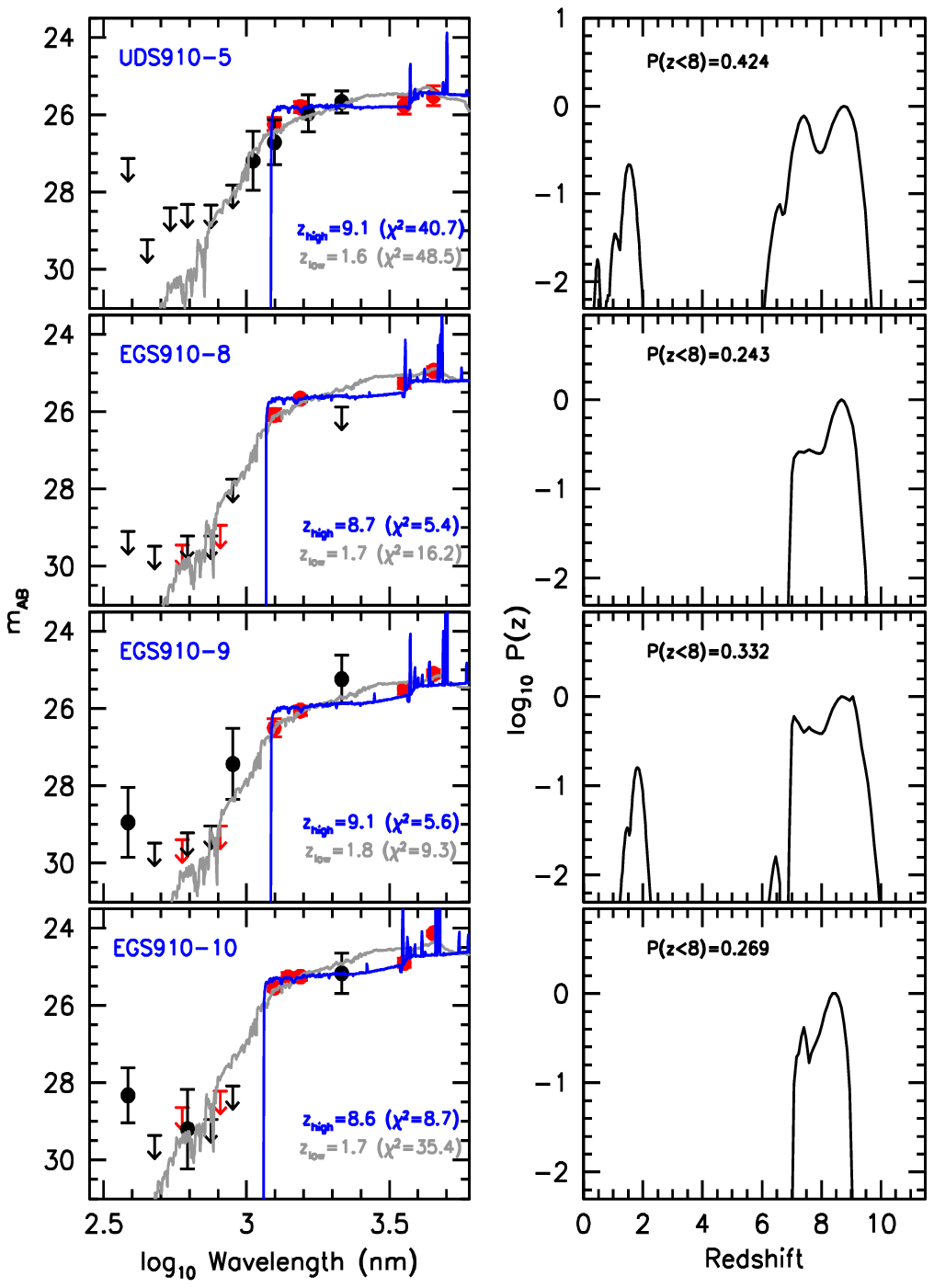}
\caption{Photometric constraints available from HST and Spitzer/IRAC
  (\textit{red circles and upper limits}) and also from various
  ground-based telescopes (\textit{black circles and upper limits}) on
  several new $z\sim9$ candidate galaxies identified within the new
  147 arcmin$^2$ search area we considered.  Upper limits are
  $1\sigma$.  The blue line shows the best-fit $z>8$ model SED we
  find, while the gray line shows the best-fit lower-redshift SED we
  derive.  The best-fit $\chi^2$ and redshifts we find for the
  candidates are also indicated in the left-most panels.  The right
  panels show the probability we compute for our $z\sim9$ candidates
  to lie at specific redshifts.  The total fractional probability that
  our $z\sim9$ candidates lie at $z<8$ is also indicated.  Our
  photometry suggests all four candidates may have redshifts in excess
  of 8.  While EGS910-10 has already been spectroscopically confirmed
  to have a redshift of 8.683 by Zitrin et al.\ (2015), the redshifts
  of the other three candidates are still quite
  uncertain.\label{fig:seds}}
\end{figure*}

\subsection{Targeted Follow-up Observations}

Also included in the present analysis are targeted follow-up
observations of two candidate $z\sim8.5$ galaxies.  The coordinates of
those candidates, COS910-5 and COS910-6, are 10:00:31.39, 02:26:39.8
and 10:00:20.12, 02:14:13.0, respectively.  1 orbit of $Y_{098}$
observations were obtained on each (GO 14459: Bouwens 2015) as part of
a cycle 23 program.

\begin{deluxetable*}{cccccc}
\tablewidth{0pt}
\tablecolumns{5}
\tabletypesize{\footnotesize}
\tablecaption{New Candidate $z\sim9$-10 Galaxies identified over the CANDELS UDS, COSMOS, and EGS programs (see \S3.2)\label{tab:catalog}}
\tablehead{
\colhead{ID} & \colhead{R.A.} & \colhead{Dec} & \colhead{$H_{160,AB}$} & \colhead{$z_{phot}$\tablenotemark{a}} & \colhead{P($z>8$)\tablenotemark{a}}}
\startdata
UDS910-5 & 02:18:03.23 & $-$05:13:21.7 & 25.8$\pm$0.1 & 9.1 & 0.58 \\
EGS910-8 & 14:20:52.51 & 53:04:11.7 & 25.7$\pm$0.1 & 8.7 & 0.76 \\
EGS910-9 & 14:20:45.23 & 53:02:01.3 & 26.1$\pm$0.1 & 9.1 & 0.67 \\
EGS910-10\tablenotemark{b} & 14:20:08.50 & 52:53:26.6 & 25.3$\pm$0.1 & 8.6\tablenotemark{c} & 0.73
\enddata
\tablenotetext{a}{Best-fit $z>4$ redshift and integrated $z>8$ likelihood for source derived from our {\it HST}+{\it Spitzer}/IRAC+ground-based photometry (\S2.1).}
\tablenotetext{b}{Also known as EGSY8p7.  This source was previously
  identified by Roberts-Borsani et al.\ (2016) as a $z\sim8.5$
  candidate and spectroscopically confirmed by Zitrin et al.\ (2015).}
\tablenotetext{c}{This source has a measured spectroscopic redshift $z=8.683$ (Zitrin et al.\ 2015).}
\end{deluxetable*}

In executing the follow-up program, we adhered to a similar strategy
as Bouwens et al.\ (2016) utilized in obtaining $Y_{105}$-band
follow-up imaging observations of their candidate $z\sim9$-10
galaxies.  The goal of the follow-up observations was to test if the
sources showed essentially no flux at $\sim$1$\mu$m and bluer
wavelengths.  This is what one would expect if they were genuine
$z\sim9$ galaxies.

Given that the sources were identified after our cycle-22 program
z9-CANDELS was complete, follow-up was requested in a subsequent HST
program (GO 14459).  Relative to the sources followed up as part of
our cycle-22 program, these sources had slightly bluer
$J_{125}-H_{160}$ colors, i.e., $<$0.5 mag, and did not satisfy the
selection criteria of Bouwens et al.\ (2016) whose
$J_{125}-H_{160}>0.5$ mag selection criteria preferentially identified
sources with redshifts $z\gtrsim 8.9$ (see Figure~\ref{fig:selc}).
Both candidates also had $H_{160}$-band magnitudes brighter than 25.5
mag and therefore could have an impact on the bright-end shape of the
$UV$ LF at $z>6$.  Given that this had been the subject of debate
(e.g., Bowler et al.\ 2014 vs. Bouwens et al.\ 2015), follow-up of
these candidates was considered to be important.

HST follow-up observations of COS910-5 and COS910-6 were obtained on
February 27, 2016 and March 1, 2016, respectively.  These observations
were reduced using our WFC3RED pipeline (Magee et al.\ 2011) and
drizzled onto the same astrometric frame as the ACS + WFC3/IR CANDELS
data described in the previous subsection.

\subsection{Photometry}

Source catalogs were constructed for the new fields using the
SExtractor software (Bertin \& Arnouts 1996) and essentially an
identical procedure to that utilized in previous papers by our team
(e.g., Bouwens et al.\ 2015, 2017).  Given that our search is for
$z\sim9$-10 galaxies, source detection is performed using the
$H_{160}$-band image alone.  Our HST color measurements are made in
smaller-scalable apertures based on a Kron (1980) factor of 1.2 and
make use of the images after PSF correction to the $H_{160}$-band PSF.
These smaller-scalable aperture flux measurements were then scaled up
to total magnitudes by first accounting for the additional flux
measured in larger scalable apertures (Kron factor of 2.5) relative to
smaller scalable apertures and second accounting for the flux outside
the larger scalable apertures and on the wings of the PSF.  The former
correction is made using the detection image, while the latter
correction is made using the tabulated encircled energies in the
WFC3/IR handbook (Dressel et al.\ 2012).

Photometry of candidate $z\sim9$-10 galaxies using the ground-based
data and Spitzer/IRAC observations can also help us constrain their
nature.  To obtain these flux measurements, we need to cope with the
very broad PSF in the ground-based and especially Spitzer/IRAC data
which results in substantial overlap between sources.  We use the
\textsc{mophongo} software (Labb{\'e} et al.\ 2010a, 2010b, 2013,
2015) to obtain flux measurements in the presence of source confusion.
\textsc{Mophongo} uses the higher resolution HST data to create
spatial templates for each source which is then used for modeling the
ground-based or Spitzer/IRAC imaging data.  The amplitudes of the
templates are varied until a good fit to the imaging data is obtained,
and then flux from the neighboring sources is subtracted.  Photometry
is performed on sources in 1.2$''$-diameter apertures for the
ground-based data and 1.8$''$-diameter apertures for the Spitzer/IRAC
data.  Finally, the results are corrected to total based on the PSFs
derived from the imaging data.

\begin{deluxetable*}{ccccccc}
\tablewidth{0pt}
\tablecolumns{7}
\tabletypesize{\footnotesize}
\tablecaption{$z\sim9$-11 Galaxy Candidates Identified over the CANDELS Fields\label{tab:catalog_conf}}
\tablehead{
\colhead{ID} & \colhead{R.A.} & \colhead{Dec} & \colhead{$H_{160,AB}$} & \colhead{$z_{phot}$\tablenotemark{b}} & \colhead{P($z>8$)} & \colhead{Ref\tablenotemark{a}}}
\startdata
\multicolumn{7}{c}{$z\sim9$ Sample}\\
\multicolumn{3}{l}{New Candidates from This Work:}\\
UDS910-5 & 02:18:03.23 & $-$05:13:21.7 & 25.8$\pm$0.1 & 9.1 & 0.58 \\
EGS910-8 & 14:20:52.51 & 53:04:11.7 & 25.7$\pm$0.1 & 8.7 & 0.76 \\
EGS910-9 & 14:20:45.23 & 53:02:01.3 & 26.1$\pm$0.1 & 9.1 & 0.67 \\
EGS910-10 & 14:20:08.50 & 52:53:26.6 & 25.3$\pm$0.1 & 8.683 & 1.0\tablenotemark{c} & [6,7]\\\\
\multicolumn{3}{l}{From Oesch et al.\ (2014) and Bouwens et al.\ (2016):}\\
COS910-1 & 10:00:30.34 & 02:23:01.6 & 26.4$\pm$0.2 & 9.0$_{-0.5}^{+0.4}$ & 0.99 & [8]\\
EGS910-0 & 14:20:23.47 & 53:01:30.5 & 26.2$\pm$0.1 & 9.1$_{-0.4}^{+0.3}$ & 0.92 & [8]\\
EGS910-3 & 14:19:45.28 & 52:54:42.5 & 26.4$\pm$0.2 & 9.0$_{-0.7}^{+0.5}$ & 0.97 & [8]\\
UDS910-1\tablenotemark{b} & 02:17:21.96 & $-$05:08:14.7 & 26.6$\pm$0.2 & 8.6$_{-0.5}^{+0.6}$ & 0.74 & [8] \\
GS-z9-1 & 03:32:32.05 & $-$27:50:41.7 & 26.6$\pm$0.2 & 9.3$\pm$0.5 & 0.9992 & [1], [8]\\
GS-z9-2  & 03:32:37.79 & $-$27:42:34.4 & 26.9$\pm$0.2 & 8.9$_{-0.3}^{+0.3}$ & 0.83 & [8]\\
GS-z9-3  & 03:32:34.99 & $-$27:49:21.6 & 26.9$\pm$0.2 & 8.8$_{-0.3}^{+0.3}$ & 0.95 & [3],[8]\\
GS-z9-4  & 03:33:07.58 & $-$27:50:55.0 & 26.8$\pm$0.1 & 8.4$_{-0.3}^{+0.2}$ & 0.97 & [3],[8]\\
GS-z9-5  & 03:32:39.96 & $-$27:42:01.9 & 26.4$\pm$0.1 & 8.7$_{-0.7}^{+0.8}$ & 0.55 & [8]\\
GN-z9-1 & 12:36:52.25 & 62:18:42.4 & 26.6$\pm$0.1 & 9.2$\pm$0.3 & $>$0.9999 & [1], [8]\\
\\
\multicolumn{7}{c}{$z\sim10$ Sample}\\
\multicolumn{3}{l}{From Oesch et al.\ (2014) and Bouwens et al.\ (2016):}\\
EGS910-2\tablenotemark{d} & 14:20:44.31 & 52:58:54.4 & 26.7$\pm$0.2 & 9.6$_{-0.5}^{+0.5}$ & 0.71 & [8]\\
GN-z10-2 & 12:37:22.74 & 62:14:22.4 & 26.8$\pm$0.1 & 9.9$\pm$0.3 & 0.9994 & [1], [2]\\
GN-z10-3 & 12:36:04.09 & 62:14:29.6 & 26.8$\pm$0.2 & 9.5$\pm$0.4 & 0.9981 & [1], [2]\\
GS-z10-1 & 03:32:26.97 & $-$27:46:28.3 & 26.9$\pm$0.2 & 9.9$\pm$0.5 & 0.9988 & [1], [2]\\
\\
\multicolumn{7}{c}{$z\sim11$ Sample}\\
\multicolumn{3}{l}{From Oesch et al.\ (2016):}\\
GN-z11 & 12:36:25.46 & 62:14:31.4 & 26.0$\pm$0.1 & 11.1$\pm$0.1 & 1.0\tablenotemark{e} & [1], [2], [4], [5]
\enddata
\tablenotetext{a}{References: [1] Oesch et al.\ 2014, [2] Bouwens et al.\ 2015, [3] McLure et al.\ 2013, [4] Oesch et al.\ 2016, [5] Bouwens et al.\ 2010, [6] Zitrin et al.\ 2015, [7] Roberts-Borsani et al.\ 2016, [8] Bouwens et al.\ 2016}
\tablenotetext{b}{The likelihood of the new candidate $z\sim9$ galaxies being secure $z>8$ sources is lower than in earlier compilations by Oesch et al.\ (2014) and Bouwens et al.\ (2016).  This is because the new candidates do not yet have deep HST coverage at $1\mu$m from $Y_{105}$-band observations as the Oesch et al.\ (2014) and Bouwens et al.\ (2016) candidates possess.}
\tablenotetext{c}{This source has a measured spectroscopic redshift $z=8.683$ (Zitrin et al.\ 2015).}
\tablenotetext{d}{The follow-up data obtained by the z9-CANDELS program did not significantly clarify the nature of this source (GO 13792: Bouwens 2014).  Nevertheless, the available observations still support this source's being a credible $z\sim9$ candidate.}
\tablenotetext{e}{This source has a measured spectroscopic redshift $z=11.1\pm0.1$ (Oesch et al.\ 2016).}
\end{deluxetable*}

\section{$z\sim9$-10 Samples}

\subsection{Selection Criteria}

In this section, we describe the selection criteria we apply to the
CANDELS UDS, COSMOS, and EGS data sets described in \S2.1 and \S2.2.
Collectively, those data sets cover an area of 601 arcmin$^2$.  As we
have emphasized earlier, 147 arcmin$^2$ of this area was left
unexplored in Bouwens et al.\ (2016).  Meanwhile, the balance of the
area, i.e., a 454 arcmin$^2$ region with the UDS, COSMOS, and EGS
regions, was re-examined using a more inclusive selection criterion
than we previously utilized.  This will allow for the identification
of more star-forming sources between $z\sim8.4$ and $z\sim9.0$ (see
Figure~\ref{fig:selc}).

In identifying probable candidate $z\sim9$-10 galaxies, we are guided
by Lyman-break-like selection criteria.  Significant spectroscopic
work has shown that the Lyman-break selection technique is very
effective in identifying galaxies at high redshifts (Steidel et
al.\ 1996; Steidel et al.\ 2003; Vanzella et al.\ 2009; Stark et
al.\ 2010; Smit et al.\ 2018).

\begin{figure}
\epsscale{1.16}
\plotone{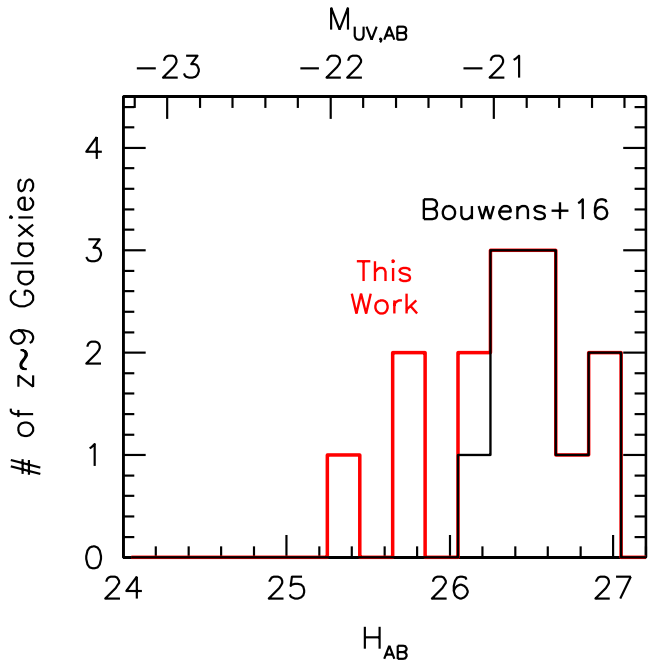}
\caption{Number of $z\sim9$ Candidates Identified over the CANDELS
  program vs. the apparent $H_{160,AB}$-band Magnitude.  The red
  histogram shows the current sample of $z\sim9$ galaxy candidates,
  while the black histogram shows sample of $z\sim9$ candidates
  identified in Bouwens et al.\ (2016).  The upper axis shows the
  corresponding absolute magnitude of galaxies at $z\sim9$ that
  corresponds to a given $H_{AB}$-band magnitude.  In the present
  selection of $z\sim9$ candidates, we find a larger fraction of
  bright ($H\leq26.1$ mag) galaxy candidates than we identified in our
  previous study.  The 25.3-mag candidate shown here was previously
  identified by Roberts-Borsani et al.\ (2016) and spectroscopically
  confirmed by Zitrin et al.\ (2015).\label{fig:numcounts}}
\end{figure}

To create a sample of potential $z\sim9$-10 candidates, we required
that sources in our selection be detected at least at 6$\sigma$ in
the $H_{160}$ band in a 0.35$''$-diameter aperture to ensure that
sources in our selection are real.  

Sources in our selection were also required to show a
$\chi_{F606W,F814W} ^2$ parameter less than 4.  Following Bouwens et
al.\ (2011), we defined the $\chi^2$ parameter as $\chi_{F606W,F814W}
^2 = \Sigma_{i=[F606W,F814W]} \textrm{SGN}(f_{i})
(f_{i}/\sigma_{i})^2$ where $f_{i}$ is the flux in band $i$ in a
consistent aperture, $\sigma_i$ is the uncertainty in this flux, and
SGN($f_{i}$) is equal to 1 if $f_{i}>0$ and $-1$ if $f_{i}<0$.

For each source that satisfied our $H_{160}$-band detection criteria
and optical non-detection criteria, we computed redshift likelihood
distributions based on our HST, Spitzer, and ground-based photometry
using the EAZY photometric redshift code (Brammer et al.\ 2008).  In
deriving the redshift likelihood distribution for each source, we make
use of EAZY\_v1.0 template set supplemented by SED templates from the
Galaxy Evolutionary Synthesis Models (GALEV: Kotulla et al.\ 2009).
Nebular continuum and emission lines were added to the later templates
using the Anders \& Fritze-v. Alvensleben (2003) prescription, a $0.2
Z_{\odot}$ metallicity, and a rest-frame EW for H$\alpha$ of 1300\AA.

Selected sources were required to have a maximum likelihood redshift
$z\geq 8.4$ following the treatment in Bouwens et al.\ (2016).  In
Bouwens et al.\ (2016), we adopted this redshift limit to provide a
midway point between our $z\sim7$ $z_{850}$ and $z\sim8$
$Y_{105}$-dropout selections where the median redshift was 6.8 and
7.9.  Use of $z=8.4$ as the lower redshift limit for $z\sim9$ samples
also allows us to slightly increase the number of sources in our
$z\sim9$ samples, providing us with more leverage on the shape of the
LF at early times.

\begin{figure*}
\epsscale{1.0}
\plotone{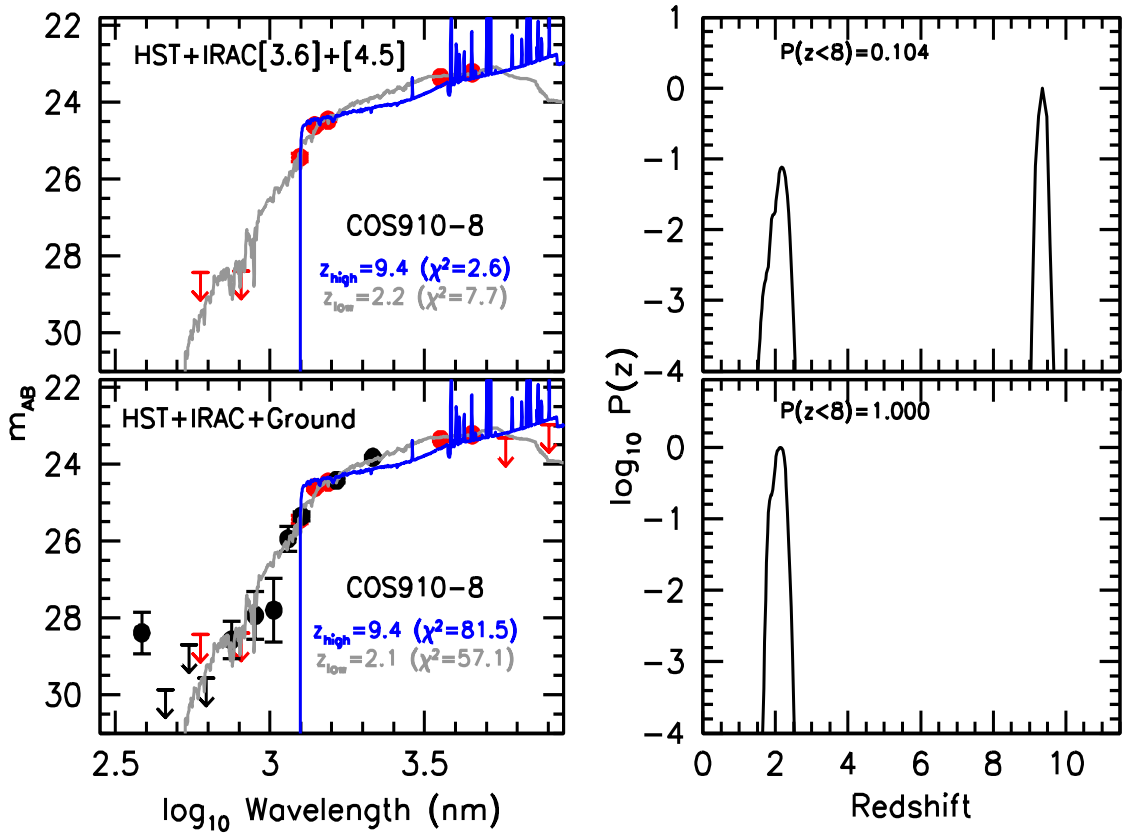}
\caption{The SED fit and redshift likelihood distribution derived for
  a source COS910-8 (10:00:34.99, 02:14:01.1) from our search fields
  that prefers a $z>9$ solution when using the HST+Spitzer/IRAC
  3.6$\mu$m+4.5$\mu$m photometry alone (\textit{upper panel}), but
  which appears to be much more likely a $z\sim2$ galaxy when
  incorporating the constraints from our ground-based photometry
  (\textit{lower panel}: see also Appendix B for other examples).  The
  lines and symbols shown in this figure are similar to
  Figure~\ref{fig:sed} from Appendix A.  In the case that we rely on
  HST+IRAC 3.6$\mu$m+4.5$\mu$m photometry alone, we estimate the
  redshift to be 9.4, with 90\% of the probability at $z>8$, but if we
  fold in the ground-based constraints, the redshift of this source
  seems much more likely to be $\sim2.1$, largely due to this source
  showing a $3\sigma$ detection in the FourStar J2 band (see
  Figure~\ref{fig:cos9108-stamps} from Appendix B) and the shape of
  continuum SED redward of the $H$ band, including constraints in the
  four IRAC bands (see Table~\ref{tab:cos9108}).  Making use of our
  ground-based photometry (where care is exercised in subtracting the
  bright neighbor) to constrain the redshift of this source, we find
  that only 2$\times$10$^{-6}$ of the integrated probability is at
  $z>8$.  If we renormalize the flux uncertainties to obtain a more
  realistic reduced $\chi^2$, the probability that COS910-8 is at
  $z>8$ is 2$\times$10$^{-4}$.\label{fig:cos9108}}
\end{figure*}

In addition to requiring that sources being at $z\geq 8.4$, we also
require that $\geq$50\% of their total integrated redshift likelihood
be at $z>8$.  For two of the $z\sim9$-10 candidates over these fields,
we could include the constraints obtained by targeted $Y_{098}$-band
observations on the sources from a cycle-23 program (\S2.3).  

Finally, the $H_{160}$$-$[3.6] colors of selected sources were
required to be bluer than 1.4 mag to avoid selecting intrinsically red
sources at lower redshifts.  With such a $H$$-$[3.6] limit, we would
identify every extreme $z\sim7$-11 source in the Bowler et
al.\ (2017), Stefanon et al.\ (2019), and Oesch et al.\ (2014)
selections over $>$1.5 deg$^2$, as we clearly show in
Figure~\ref{fig:h36} from Appendix B.  Conversely, sources with
$H-[3.6]$ colors redder than 1.4 mag almost always appear to be at
$z<4$ (for the cases we have examined in Appendix B).

The above criteria differ from those utilized in Bouwens et
al.\ (2015, 2016) in that they allow for the selection of sources with
$J_{125}-H_{160}$ colors bluer than 0.5 mag.  Note that a
$J_{125}-H_{160}$ color of 0.5 mag corresponds to a redshift $z$ of
$\sim$8.9 for sources with $UV$-continuum slopes of $-2$ (see
Figure~\ref{fig:selc}).  Bouwens et al.\ (2016) explicitly did not
consider such sources to focus on the highest redshift sources over
CANDELS with their follow-up observations.

\subsection{$z\sim9$-10 Sample}

Application of the above selection criteria to our photometric
catalogs allowed us to identify three new $z\sim9$ candidates over the
previously unexplored area that we searched from CANDELS UDS, COSMOS,
and EGS (a $\sim$147 arcmin$^2$ area).  One of these candidates was
identified over the CANDELS UDS area, while the other two candidates
were from the new CANDELS EGS area we examined.  No new $z\sim9$
candidates were identified over the area we considered from CANDELS
COSMOS.

A fourth $z\sim9$ galaxy was identified over the original 454
arcmin$^2$ area previously considered by us in Bouwens et al.\ (2016)
over CANDELS UDS, COSMOS, and EGS.  This source is the now well-known
$z=8.683\pm0.003$ galaxy initially identified by Roberts-Borsani et
al.\ (2016) and spectroscopically confirmed by Zitrin et al.\ (2015).

Only one of the two $z\sim9$ candidates targeted with HST
$Y_{098}$-band observations from our cycle 23 program was determined
to have a probable redshift in excess of 8.  However, even in the case
of that source, our best estimate for its redshift is 8.3, which would
put it below our selection limit of 8.4.

Sources in the selected sample had $H_{160,AB}$ band magnitudes
between 25.3 and 26.1 mag, similar to those identified by Bouwens et
al.\ (2016) but roughly $\sim$0.5 mag brighter in terms of their
overall magnitude.  See Figure~\ref{fig:numcounts}.  The brightest
source in our selection was previously identified by Roberts-Borsani
et al.\ (2016) by focusing on the brightest ($H_{160,AB}<25.5$)
sources and requiring them to be undetected in the optical data while
showing very red [3.6]$-$[4.5] colors, as is expected at $z>7$ due to
the contribution of [OIII]+H$\beta$ line emission to the 4.5$\mu$m
fluxes.
 
In addition to the three new $z\sim9$ candidate galaxies identified
here, there are also 28 other $z\geq 7$ candidate galaxies identified
over the CANDELS UDS, COSMOS, and EGS fields which, while mostly being
lower quality candidates in general, could nevertheless be at
$z\sim9$-10 in a few cases.  These sources are presented in Appendix
A, B, and in Tables~\ref{tab:catalog_z80} and \ref{tab:catalog_lq}.
Three of these candidates were identified as part of our earlier
$z\sim9$-10 search (Bouwens et al.\ 2016: their Table 7), but which we
have been unable to thus far confirm through the acquisition of deeper
HST data.  6 of the candidates have redshifts $z$ $\leq$ 8.4.  16 of
the candidates were identified using the HST + Spitzer/IRAC
3.6$\mu$m+4.5$\mu$m data alone, in case confusion in the ground-based
observations affected our selection (e.g., if there are spurious
detections in the ground-based optical data).

Of all the candidate $z\sim9$-10 sources included in Appendix B, a
particularly interesting case is COS910-8, given its exceptional
brightness $H_{160,AB}\sim24.5$ mag.  Using the HST+Spitzer/IRAC
3.6$\mu$m+4.5$\mu$m photometry, we estimate a redshift of $z\sim9.4$
for the source, with 90\% of the probability lying at $z>8$.  While it
would appear to be quite a compelling candidate based on our
HST+Spitzer/IRAC 3.6$\mu$m+4.5$\mu$m photometry, our assessment of
this candidate depends sensitively on whether we incorporate
constraints from our ground-based photometry or not.  While we
initially considered the possibility that this source might have been
erroneously excluded from our earlier $z\sim9$-10 selections (Bouwens
et al.\ 2016) due to inclusion of optical flux from a bright neighbor,
photometry on the source, after careful subtraction of the neighbor,
indicates that this source is much more likely at $z\sim2.1$,
particularly owing to the apparent detection of this source in the
$\sim$1.15$\mu$m J2 band at $3\sigma$ (see
Figure~\ref{fig:cos9108-stamps} in Appendix B), the SED shape of the
source defined by the $H$, $K_s$, and Spitzer/IRAC flux measurements,
and tentative 1-2$\sigma$ detections of the source in the $i$, $z$, and
$Y$ bands.  On the basis of the computed $\chi^2$, the probability
that COS910-8 is at $z>8$ is 2$\times$10$^{-6}$.  Given the high value
of $\chi_{min} ^2$ relative to the number of constraints minus fit
degrees of freedom, it is possible that our estimated flux
uncertainties are too small.  If we renormalize these uncertainties so
that the reduced $\chi^2$ is 1 for the best-fit solution, the
probability that this candidate is at $z>8$ is 2$\times$10$^{-4}$.

We now return our discussion to the new $z\sim9$ $P(z>8)>0.5$ candidates
identified as part of this study (Table~\ref{tab:catalog}).  Each of
these candidates show a $>$50\% probability of having a redshift $z$
$>$ 8.  Nevertheless, each of them could have a redshift $z\lesssim
8$.  Previously, Bouwens et al.\ (2016) had identified 9 $z\sim9$-10
candidates that showed a $\gtrsim$58\% probability of lying at $z>8$
and followed them up with $Y_{105}$-band observations to test their
robustess.  Based on the follow-up observations, four sources were
confirmed as robust $z>8$ candidates, four sources were found to
prefer a redshift $z\lesssim8$, and for one source, the follow-up
observations were not helpful in clarifying the nature of the source.

We will assume that follow-up of the new $z\sim9$ candidates with
$Y_{105}$-band observations would yield a similar $\sim$50\%
contamination fraction to that obtained in our earlier follow-up
efforts.  Given that one of the four sources in our expanded selection
was already spectroscopically confirmed, i.e., the Zitrin et
al.\ (2015) $z=8.683$ source, we assume that 2.5 of the $z\sim9$
candidates are bona-fide $z\sim9$ galaxies and 1.5 of our $z\sim9$
candidate have redshifts $z<8$.

We will combine the new identifications of $z\sim9$ candidate galaxies
with the Oesch et al.\ (2014) and Bouwens et al.\ (2016)
identifications of $z\sim9$-10 candidate galaxies.
Table~\ref{tab:catalog_conf} provides a comprehensive list of all the
$z\sim9$-10 candidates we have identified.

In total, our selection includes 9 high-confidence ($P(z>8)>0.8$)
$z\sim9$ candidate galaxies, 3 high-confidence $z\sim10$ candidate
galaxies, and 1 spectroscopically confirmed $z\sim11$ galaxy.  Our
$z\sim9$ and $z\sim10$ selections include 5 sources and 1 source,
respectively, with $>$55\% of the probability lying at $z>8$.  If all
of these sources lie at $z>8$, we find a total of 19 $z\sim9$-11
sources in the $\sim$883 arcmin$^2$ area that make up the 5 CANDELS
fields.  This translates to a surface density of 1 bright $z\sim9$-11
candidate per 47 arcmin$^2$ (i.e., $\approx$10 WFC3/IR pointings).

\begin{figure*}
\epsscale{1.17}
\plotone{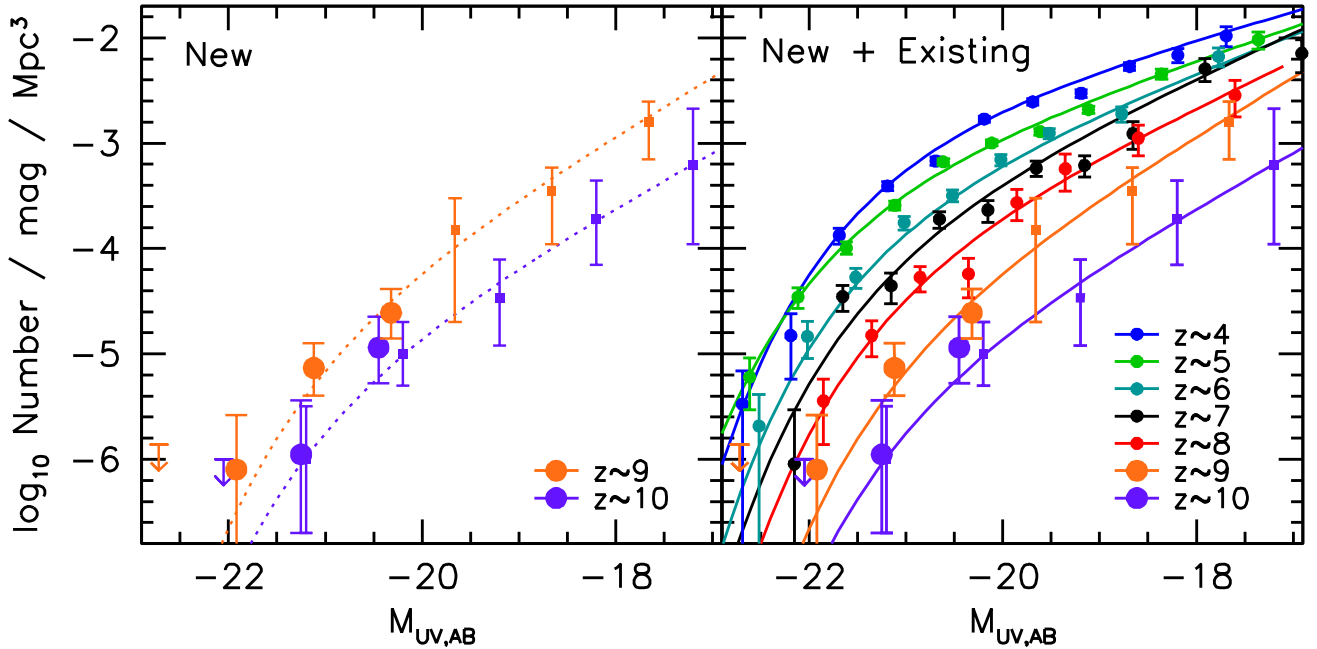}
\caption{(\textit{left}) Current determinations of the stepwise $UV$
  LF at $z\sim9$ and $z\sim10$ (\textit{large solid orange and purple
    circles, respectively}: \S4.1).  The plotted error bars are
  $1\sigma$.  In deriving the $z\sim10$ LF, sources out to $z\sim11$
  (i.e., Oesch et al.\ 2016) have been included.  Table~\ref{tab:swlf}
  presents our new results in tabular form.  The dotted orange and
  purple lines show the $z\sim9$ and $z\sim10$ LF determinations from
  Bouwens et al.\ (2019, in prep) and Oesch et al.\ (2018),
  respectively.  For context, the $z\sim9$ LF from Oesch et
  al.\ (2013: \textit{orange solid squares}) and $z\sim10$ LF from
  Oesch et al.\ (2018: (\textit{purple solid squares}) at lower
  luminosities are also shown.  (\textit{right}) Our new LF
  determinations at $z\sim9$ and $z\sim10$ are shown alongside of
  those determinations by Bouwens et al.\ (2015) at $z\sim4$
  (\textit{blue solid circles}), $z\sim5$ (\textit{green solid
    circles}), $z\sim6$ (\textit{cyan solid circles}), $z\sim7$
  (\textit{black solid circles}), and $z\sim8$ (\textit{red solid
    circles}).  The $z\sim9$ and $z\sim10$ results from Bouwens et
  al.\ (2019, in prep), Oesch et al.\ (2013), and Oesch et al.\ (2018)
  from the left panel are also shown in the right panel.  The
  Schechter function fits derived at $z\sim4$, $z\sim5$, $z\sim6$,
  $z\sim7$, and $z\sim8$ by Bouwens et al.\ (2015) are shown with the
  blue, green, cyan, black, red, orange, and purple lines,
  respectively.\label{fig:lfall}}
\end{figure*}

\section{Luminosity Function Results}

In the present study, we have expanded our search for $z\sim9$-10
galaxies to cover a $\sim$883 arcmin$^2$ area within CANDELS, which is
an improvement on the 736 arcmin$^2$ area we previously considered in
Bouwens et al.\ (2016).  Thanks to our expanded search area, we were
able to expand our $z\sim9$-10 selection from 15 candidate $z\sim9$-10
galaxies over the CANDELS fields to 19 such candidates.

In this section, we make use of our expanded $z\sim9$-10 sample and
search area to improve our constraints on the prevalence of bright
galaxies at $z\sim9$-10.

\begin{deluxetable}{lc}
\tablewidth{0pt}
\tabletypesize{\footnotesize}
\tablecaption{New Stepwise Determinations of the $UV$ LFs at $z\sim9$ and $z\sim10$ using a $\sim$883 arcmin$^2$ search area over all 5 CANDELS fields\label{tab:swlf}}
\tablehead{
\colhead{$M_{UV,AB}$\tablenotemark{a}} & \colhead{$\phi_k$ (10$^{-3}$ Mpc$^{-3}$ mag$^{-1}$)}}
\startdata
\multicolumn{2}{c}{$z\sim9$ galaxies}\\
$-$22.72 & $<$0.0014\tablenotemark{b}\\
$-$21.92 & 0.0008$_{-0.0007}^{+0.0018}$\\
$-$21.12 & 0.0074$_{-0.0034}^{+0.0053}$\\
$-$20.32 & 0.0246$_{-0.0106}^{+0.0166}$\\
\multicolumn{2}{c}{$z\sim10$ galaxies}\\
$-$22.84 & $<$0.0014\tablenotemark{b}\\
$-$22.05 & $<$0.0010\tablenotemark{b}\\
$-$21.25 & 0.0011$_{-0.0009}^{+0.0025}$\\
$-$20.45 & 0.0115$_{-0.0062}^{+0.0111}$
\enddata 
\tablenotetext{a}{Derived at a rest-frame wavelength of 1600\AA.}
\tablenotetext{b}{$1\sigma$ upper limit.}
\end{deluxetable}

\subsection{Luminosity Function Results}

As in other studies of the LF (e.g., Steidel et al.\ 1999; Bouwens et
al.\ 2007, 2008), we achieve constraints on our model LFs by comparing
in detail with the surface density of $z\sim9$-10 candidates found in
the data sets.

Given the small number of $z\sim9$-10 candidates in our bright samples
and especially per search field, we use luminosity bins of width 0.8
mag in constructing a $UV$ LF and model the counts in each
observational bin in the data sets as Poissonian.  This results in the
following estimated likelihood ${\cal L}$ for a model LF given a set
of observations:
\begin{equation}
{\cal L}=\Pi_{i,j} e^{-N_{exp,i,j}} \frac{(N_{exp,i,j})^{N_{obs,i,j}}}{(N_{obs,i,j})!}
\label{eq:ml}
\end{equation}
where $\Pi_{i,j}$ is the product symbol and which runs over all search
fields each denoted by index $i$ and over all magnitude intervals
denoted by index $j$, where $N_{obs,i,j}$ is the observed number of
sources in search field $i$ and magnitude interval $j$, and where
$N_{exp,i,j}$ is the expected number of sources in search field $i$
and magnitude interval $j$.

We compute the expected number of sources per bin, i.e., $N_m$, from a
model LF as follows: 
\begin{equation}
N_m = \Sigma _{k} \phi_k V_{m,k}
\label{eq:numcountg}
\end{equation}
where $N_m$ is the surface density of galaxies in some search field
with magnitude $m$, $\phi_k$ is the volume density of galaxies with
absolute magnitude $k$, and $V_{m,k}$ is the effective selection
volume for which galaxies with absolute magnitude $k$ will both
satisfy our dropout selection criteria and be observed to have an
apparent magnitude $m$.  The binning we adopt for $N_m$ and $V_{m,k}$
is the same 0.8-mag binning as we adopt for the stepwise LF $\phi_k$.

\begin{figure*}
\epsscale{0.64}
\plotone{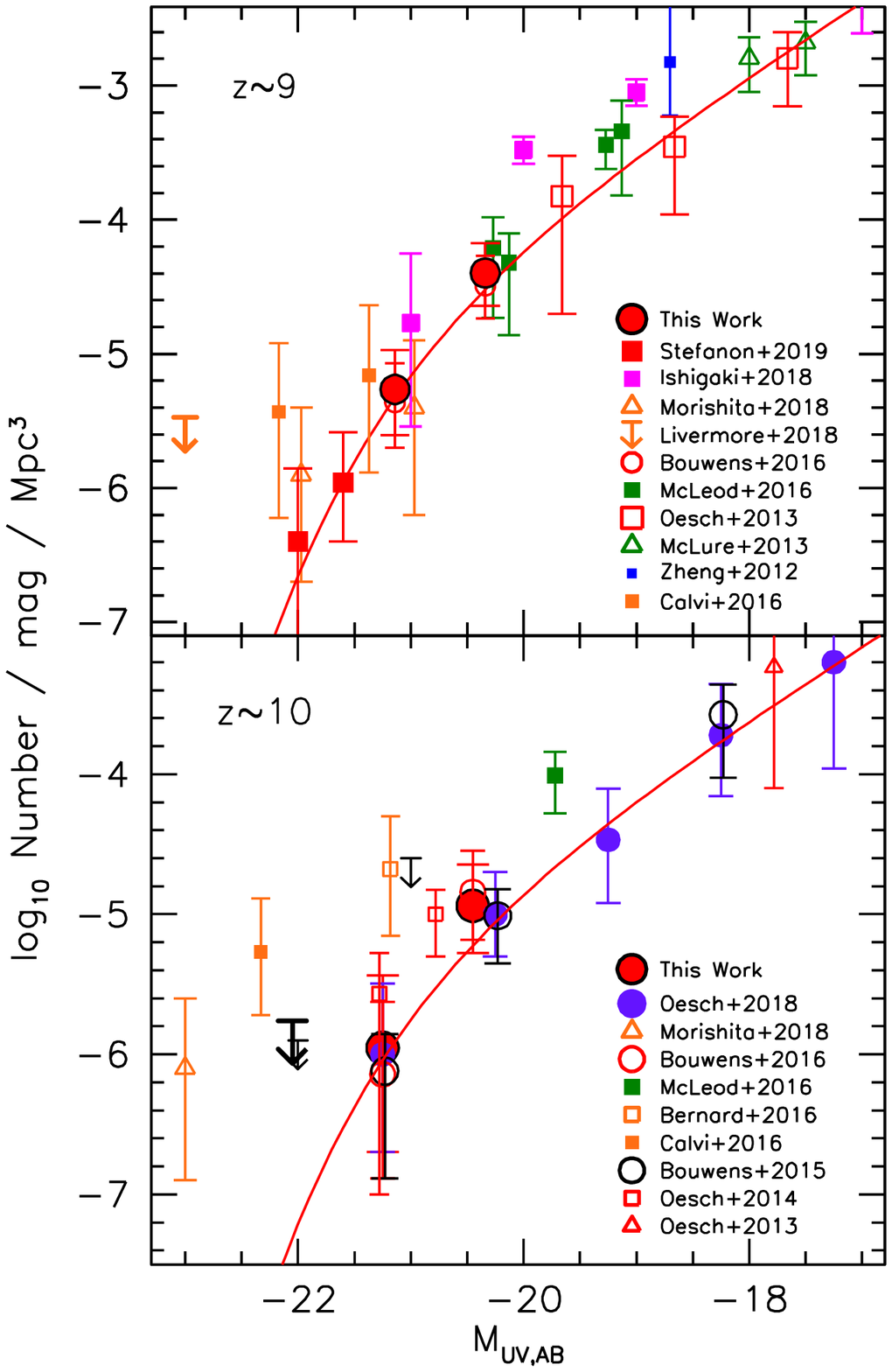}
\caption{Comparison of our $z\sim9$ and $z\sim10$ determinations
  incorporating the present larger search area over CANDELS
  (\textit{upper and lower panels}), with noteworthy previous
  determinations of the LF in these same redshift intervals (Zheng et
  al.\ 2012; McLure et al.\ 2013; Oesch et al.\ 2013, 2014; Bouwens et
  al.\ 2015, 2016; McLeod et al.\ 2016; Oesch et al.\ 2018; Morishita
  et al.\ 2018; Livermore et al.\ 2018; Ishigaki et al.\ 2018;
  Stefanon et al.\ 2019: \S4.2).  Two $z\sim9$ determinations from
  McLeod et al.\ (2016) are shown, indicating both their blank-field
  and lensing field estimates.  The red lines show the $z\sim9$ and
  $z\sim10$ LF determinations from Bouwens et al.\ (2019, in prep) and
  Oesch et al.\ (2018), respectively.\label{fig:complf}}
\end{figure*}

Our procedure for estimating the selection volume $V_{m,k}$ is almost
identical to what we do in many other of our recent papers on the
high-redshift luminosity function (e.g., Bouwens et al.\ 2010, 2011,
2015).  We begin by first constructing a mock catalog of sources with
redshifts lying between $z\sim7.5$ to $z\sim11.5$, $UV$ luminosities
lying between $-$23 and $-$19.5 mag, and with random spatial positions
within the CANDELS UDS, COSMOS, and EGS fields.  Realistic images for
sources from our mock catalogs are first simulated by artificially
redshifting similar luminosity $z\sim4$ sources from the Bouwens et
al.\ (2015) HUDF samples to $z\sim9$-11 using the Bouwens et
al.\ (1998, 2003) cloning procedure.  Source sizes are scaled as
$(1+z)^{-1.1}$ to match the galaxy size vs. redshift trend observed to
$z\sim9$-10 (e.g., Ono et al.\ 2013; Holwerda et al.\ 2015; Shibuya et
al.\ 2015).  The $UV$-continuum slopes are set equal to $-1.8$,
consistent with that measured at high luminosities at $z\sim5$-8
(Bouwens et al.\ 2012, 2014; Finkelstein et al.\ 2012; Willott et
al.\ 2013; Rogers et al.\ 2014), with the dispersion set to 0.3
(Bouwens et al.\ 2012; Castellano et al.\ 2012).

In addition to producing HST images for sources in our mock catalogs,
ground-based and Spitzer/IRAC images are also generated for all
sources in our catalogs.  These images are created by convolving the
HST images by the appropriate PSF-matching kernel, i.e.,
HST-to-Spitzer/IRAC or HST-to-ground-based-image.  The simulated HST,
ground-based, and Spitzer/IRAC images of our mock sources are added to
the real data and sources are detected, selected, and characterized in
the same way as sources in the real observations.  In this way, we
compute the selection volume $V_{m,k}$, where sources in the absolute
magnitude interval $k$ are selected and found to have an apparent
magnitude in the interval $m$.

We combine our new $z\sim9$ candidate galaxies with those previously
identified over the CANDELS fields.  Motivated by the results of
Bouwens et al.\ (2016: see \S3.2) who only are able to confirm 50\% of
the $z>8.4$ candidates with the $Y_{105}$-band follow-up observations,
we assume the same for our new candidates lacking follow-up
$Y_{105}$-band observations.  For two $z\sim9$ and $z\sim10$
candidates for which our cycle-22 follow-up observations were
indeterminant, i.e., UDS910-1 and EGS910-2, we treat these sources as
0.5 $z\sim9$ and $z\sim10$ candidates, consistent with the 50\%
confirmation rate achieved with follow-up observations and consistent
with the procedure applied in Bouwens et al.\ (2016).  We treat all of
the other previously presented candidates from Bouwens et al.\ (2016)
as full candidates, with the exception of GS-z9-5 which we treat as
half a candidate.

As discussed in Roberts-Borsani et al.\ (2016), there is reason to
believe that the bright $z=8.683$ source EGS910-10 (or EGSY8p7 as
Zitrin et al.\ 2015) may benefit from lensing magnification from two
massive intermediate redshift galaxies that lie within $3''$ of it.
While the degree of magnification is uncertain, we assume it is the
same factor of $\approx$2 that Roberts-Borsani et al.\ (2016)
estimate, and therefore shift the source in magnitude by 0.75 mag.
Since $z\sim11$ galaxies also satisfy our $z\sim10$ selection criteria
(but likely constitute a very small fraction of that sample), we
include the Oesch et al.\ (2016) GN-z11 $z=11.1\pm0.1$ source in our
$z\sim10$ sample.

Using our expanded $z\sim9$ and $z\sim10$ samples and computed
volumes, we recomputed the stepwise rest-$UV$ LFs at $z\sim9$ and
$z\sim10$.  Our results are presented in Table~\ref{tab:swlf} and the
left panel of Figure~\ref{fig:lfall}.  Our results are shown in the
context of our previous results at $z\sim4$, $z\sim5$, $z\sim6$,
$z\sim7$, and $z\sim8$ in the right panel to Figure~\ref{fig:lfall},
and there is reasonably smooth evolution with redshift.

We can quantify the redshift evolution better by using our new
stepwise $UV$ LF determination to compute the total luminosity density
brightward of $-20$ mag.  The integrated luminosity density brightward
of $-20$ is $10^{24.23_{-0.30}^{+0.11}}$ ergs/s/Hz/Mpc$^3$ at $z\sim9$
and $10^{23.89_{-0.97}^{+0.16}}$ ergs/s/Hz/Mpc$^3$ at $z\sim10$.  We
can compare these luminosity densities with that seen at $z\sim8$ to
the same limiting magnitude, we find $10^{24.91_{-0.06}^{+0.06}}$,
using the $z\sim8$ LF from Bouwens et al.\ (2015) LF.

As in previous work (e.g., Oesch et al.\ 2014, 2018), the luminosity
density we find at $z\sim9$, and especially at $z\sim10$, is much
smaller than found at $z\sim8$, just 100-200 Myr later in cosmic time.
The evolution from $z\sim10$ to $z\sim8$ is a factor of 10, consistent
with estimates from much previous work (e.g., Oesch et al.\ 2012;
Ellis et al.\ 2013; Bouwens et al.\ 2015; Oesch et al.\ 2018; Ishigaki
et al.\ 2018).  The present results are therefore broadly consistent
with a rapid evolution in the $UV$ LF results at $z>8$, as earlier
suggested by the ``accelerated'' evolution scenario of Oesch et
al.\ (2012).  Given the small amount of dust extinction that appear to
be present in most galaxies at $z\geq 8$ (e.g., Bouwens et al.\ 2014,
2016), a factor of ten evolution in the luminosity density is
equivalent to the star formation rate density evolving by a factor of
10.

\subsection{Comparison to Previous $z\sim9$-10 Search Results}

It is useful for us to compare our new $z\sim9$ and $z\sim10$ LF
constraints from CANDELS with the substantial number of previous
determinations of these LFs (Zheng et al.\ 2012; McLure et al.\ 2013;
Oesch et al.\ 2013, 2014; Bouwens et al.\ 2014b, 2015, 2016; Calvi et
al.\ 2016; McLeod et al.\ 2016; Bernard et al.\ 2016; Morishita
al.\ 2018; Livermore et al.\ 2018; Stefanon et al.\ 2019).  A
comparison of our new results with earlier results is presented in
Figure~\ref{fig:complf}.  Both the blank-field and the lensing field
$z\sim9$ LF determinations from McLeod et al.\ (2016) are shown with
the green squares.\footnote{For simplicity, only the NFW lensing
  results from McLeod et al.\ (2016) are presented from the CLASH
  (Postman et al.\ 2012) and HFF programs.}  The red lines in these
figures show the $z\sim9$ and $z\sim10$ LF determinations from Bouwens
et al.\ (2019, in prep) and Oesch et al.\ (2018), respectively.  The
Bouwens et al.\ (2019, in prep) $z\sim9$ LF determination leverages
both the present $z\sim9$ search results and those from the Hubble
Frontier Field parallels, the HUDF, HUDF09-1, and HUDF09-2.

As the figure illustrates, our results are in reasonable agreement
with most previous studies at $z\sim9$.  More scatter is seen in
results at $z\sim10$.  Relative to the Bouwens et al.\ (2016) $z\sim9$
LF determinations, we infer a $\approx2\times$ higher volume density
of bright $M_{UV,AB}\leq-21.1$ galaxies.  This is directly a
consequence of the larger number of bright $z\sim9$ galaxies
identified here (Figure~\ref{fig:numcounts}).  At $z\sim9$, only the
Ishigaki et al.\ (2018) $-20$-mag and $-$19 mag points lie
significantly in excess of the median volume density trend defined by
the many different determinations of the LF, i.e., by factors of
$\sim$6 and $\sim$3.

At $z\sim10$, there are significant differences between the volume
density probes from pure parallel BoRG/HIPPIES observations, e.g.,
Calvi et al.\ (2015), Bernard et al.\ (2016), and Morishita et
al.\ (2018), and those obtained from legacy fields like CANDELS, i.e.,
Bouwens et al.\ (2015, 2016) and Oesch et al.\ (2018).  One concern
with the $z\sim10$ candidates identified from the pure-parallel
programs is contamination from lower redshift candidates.  Not
surprisingly, pure-parallel studies like Morishita et al.\ (2018) --
who make use of Spitzer/IRAC data to eliminate lower-redshift
contaminants from high-redshift samples -- are more in line with
results from fields like CANDELS where such multi-wavelength data are
available to discriminate against such contaminants.

\subsection{Field-to-Field Variations}

Finally, with our new expanded sample of bright $z\sim9$ candidates
over the CANDELS fields, we investigate possible field-to-field
variations in the volume density of bright sources.  We do so, by
treating each CANDELS field as independent and using the observed
sample from each field to derive the normalization of the $UV$ LF.

\begin{figure}
\epsscale{1.16} \plotone{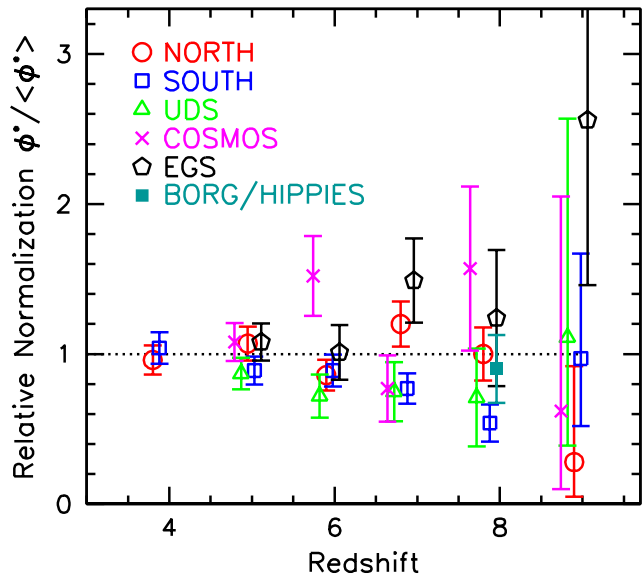}
\caption{Relative normalization inferred for the $z\sim9$ LF in each
  of the five CANDELS fields, including GOODS-North (\textit{red
    circles}), GOODS-South (\textit{blue squares}), UDS (\textit{green
    triangles}), COSMOS (\textit{magenta crosses}), and EGS
  (\textit{black pentagons}: see \S4.3).  Both the maximum likelihood
  determination (\textit{solid points}) and the $1\sigma$ uncertainty
  (\textit{error bars}) are shown.  Also shown are estimates of the
  relative normalization of $z\sim4$, 5, 6, 7, and 8 LFs in each of
  the same five CANDELS fields, as well as the BoRG/HIPPIES
  pure-parallel fields at $z\sim8$ (\textit{solid cyan square}), as
  derived in Bouwens et al.\ (2015).\label{fig:f2f}}
\end{figure}

For our LF fits, we use the same formalism as described in \S4.1, but
instead using a Schechter function to create an equivalent binned LF
in 0.1 mag intervals.  For simplicity, we fix $M^*$ and $\alpha$ to
$-21.05$ and $-2.34$, respectively, the values we derive in Bouwens et
al.\ (2019, in prep) from a comprehensive analysis of the LF from
$z\sim2$ to $z\sim9$, including the HUDF, HUDF parallel fields, HFF
parallel fields, the ERS fields, all five CANDELS fields, and 50
pure-parallel fields.

We determine the best-fit $\phi^*$ for each CANDELS field and then
compute a relative normalization $\phi^*/<\phi^*>$ by comparing the
normalization derived for a single CANDELS field with that derived in
Bouwens et al.\ (2019, in prep) considering all CANDELS fields
together.  Each of the new $z\sim9$ candidates identified here are
treated as 0.5 $z\sim9$ sources to account for the possibility that
each may lie at $z<8$ (which is identical to our treatment of these
candidates for our primary LF determinations [\S4.1]).  We also note
that our calculation of the normalization factors for each field
requires that we account for both the differing depths and areas of
each CANDELS field (as increases in either quantify would increase the
total expected number of sources).

The results are presented in Figure~\ref{fig:f2f}.  Uncertainties are
computed based on the Poissonian uncertainties given the small number
of $z\sim9$ galaxies per CANDELS field.  The relative normalizations
found by Bouwens et al.\ (2015) for star-forming galaxies at
$z\sim4$-8 in each of the five CANDELS fields are also shown as a
function of redshift.

It is interesting to compute the scatter in the relative normalization
and compare it to what is expected from simple Poissonian variations.
The RMS scatter in the relative normalizations is 0.67.  If we create
10$^5$ realizations of the CANDELS fields according to the expected
number of $z\sim9$ galaxies per CANDELS field (given their differing
depths and areas), the median RMS scatter we compute is 0.56.  

While the RMS scatter in the observations is larger than that seen for
our median Monte-Carlo simulation, 30\% of these simulations give an
RMS scatter of similar size or larger than the observed value 0.67.
As such, even models with no field-to-field variations are consistent
with our observational results.

In summary, we have tried to quantify how the normalization of the
$z\sim9$ $UV$ LF varies from one CANDELS field to another.
Unfortunately, the number of $z\sim9$ candidates per CANDELS field is
not sufficiently large to determine this accurately with present data
sets.

\section{Summary}

In this paper, we present new constraints on the bright end of the
$z\sim9$ and $z\sim10$ LFs based on a search $z\sim9$-10 candidate
galaxies within a $\sim$883 arcmin$^2$ area over the five CANDELS
fields.  The present search includes a 601 arcmin$^2$ area over the
CANDELS-WIDE UDS, COSMOS, and EGS fields.  

The present selection expands on our previous selection of $z\sim9$-10
galaxies over these same CANDELS fields (Bouwens et al.\ 2016) to
include an additional $\sim$147 arcmin$^2$ in search area.  We were
able to add to our overall selection area within CANDELS by
considering those regions which, while having deep WFC3/IR data, did
not have deep ACS optical data available from the CANDELS program.

The present selection also considered sources with a broader range of
$J_{125}-H_{160}$ colors in our identification of $z\sim9$-10
candidate galaxies than in our previous study.  Full utilization of
the Spitzer/IRAC observations from S-CANDELS (Ashby et al.\ 2016) and
the ground-based optical and near-IR observations is made to refine
our selection.

In total, we used the present larger search area to identify three new
$z\sim9$-10 candidate galaxies.  None of these sources were present in
our earlier Bouwens et al.\ (2015, 2016) catalogs or any other
published catalogs in the literature.  We also identified a fourth
candidate with our inclusive selection criteria, which while not
identified specifically in our $z\sim9$-10 searches, was identified by
Roberts-Borsani et al.\ (2016) using an IRAC [3.6]$-$[4.5]$>$0.5
selection designed to pick out bright galaxies at $z>7$, and which has
already been spectroscopically confirmed to lie at $z=8.68$ (Zitrin et
al.\ 2015).

In creating our expanded $z\sim9$ samples, we also make use of
additional follow-up observations obtained with HST in the
$Y_{098}$-band (GO 14459: Bouwens 2015) of two bright ($H<25.5$),
candidate $z\sim9$ galaxies identified over the CANDELS fields
(Appendix A).  These candidates had bluer $J_{125}-H_{160}$ colors
than the $>$0.5 mag limit we had previously considered.  While one
candidate is not confirmed to have a redshift of $z>8$, being well
detected in the HST $Y_{098}$-band data, the other candidate is
confirmed to lie at $z>8$, but with a redshift of 8.3 -- too low for
inclusion in our $z\sim9$ selection.

Adding our newly identified $z\sim9$ candidates to our previous
samples (from Bouwens et al.\ 2016), we identify a total sample of 14
bright $z\sim9$ galaxy candidates over a $\sim$883 arcmin$^2$ area in
CANDELS.  5 candidate $z\sim10$-11 galaxies are found in the same area
in CANDELS.  This is equivalent to identifying 1 $z\sim9$-11 candidate
per 47 arcmin$^2$ ($\approx$10 WFC3/IR fields).  Interestingly, our
expanded selection of $z\sim9$ galaxies has $UV$ luminosities which
are generally brighter (by 0.1 to 0.4 mag) than in our previous
selection of $z\sim9$ galaxies (compiled in Bouwens et al.\ 2016).

In addition to the 19 candidate $z\sim9$-10 galaxies we identify over
CANDELS that make up our main selection, we also identify 28 mostly
lower likelihood candidates (Appendix B).  During this process, we
consider sources selected from the HST + Spitzer/IRAC
3.6$\mu$m+4.5$\mu$m data alone in case confusion in the ground-based
results in some incompleteness.  While a few of those candidates
appear to be reliable based on those data, addition of the
ground-based constraints show that many are much more likely to be at
$z<4$ (see Figure~\ref{fig:cos9108} and \ref{fig:cos9108-stamps} from
Appendix B).  In Appendix C, Keck observations are used to improve our
constraints on the nature of a candidate examined in Appendix B.  The
entire discussion provided in the Appendices is useful in illustrating
the challenges present in selecting a high-quality sample of $z>8$
galaxies based on current data sets.

We use this expanded selection of $z\sim9$-10 candidate galaxies to
refine our determinstion of the high-luminosity end of the $UV$ LF at
$z\sim9$ and $z\sim10$.  Our revised determinations show a
$\approx2$$\times$ higher volume density of bright ($M_{UV,AB}\leq
-21.1$) $z\sim9$ galaxies than found by Bouwens et al.\ (2016).  This
owes to the increased fraction of bright ($m_{AB}\leq26.1$) $z\sim9$
galaxies identified in the new area we probe
(Figure~\ref{fig:numcounts}).

By comparing the number of bright $z\sim9$ galaxies identified with
the number expected, we attempted to estimate the relative volume
density of $z\sim9$ galaxies per CANDELS field.  The RMS variation we
found was in excess of that expected from Poissonian statistics.
Nevertheless, we found that the observed scatter was not especially
significant and we could reproduce it, adopting simple Poissonian
statistics, in as many as $\sim$30\% of our Monte-Carlo trials.  To
quantify this better, clearly deeper observations are required over
all 5 CANDELS fields to identify a larger number of $z\sim9$ galaxies.

With our new results, we confirm the strong evolution seen in the $UV$
LF at $z>8$ in previous work (Oesch et al.\ 2012, 2014; Bouwens et
al.\ 2015, 2016; Oesch et al.\ 2018), with a factor of 10 evolution
from $z\sim10$ to $z\sim8$.  The present results are broadly
consistent with the ``accelerated'' evolution scenario suggested by
Oesch et al.\ (2012).

Better constraints on the volume density of bright $z\sim9$ galaxies
could be obtained by continuing our exploitation of wide-area VISTA
surveys as have been conducted by Bowler et al.\ (2014), Stefanon et
al.\ (2017), and Stefanon et al.\ (2019), by surveying much wider area
fields to faint magnitudes with HST, by improving our exploitation of
the archival and pure parallel HST + Spitzer/IRAC data (e.g.,
Morishita et al.\ 2018), and in the future with JWST, Euclid, and
WFIRST.

\acknowledgements

The feedback on our analysis from an anonymous expert referee
significantly improved our manuscript.  This work is based on
observations taken by the CANDELS Multi-Cycle Treasury Program, the
3D-HST Treasury Program (GO 12177 and 12328), and the z9-CANDELS
program (GO 13792) with the NASA/ESA HST, which is operated by the
Association of Universities for Research in Astronomy, Inc., under
NASA contract NAS5-26555.  This work also relies on observations made
with the Spitzer Space Telescope, which is operated by the Jet
Propulsion Laboratory, California Institute of Technology under a
contract with NASA. Support for this work was provided by NASA through
an award issued by JPL/Caltech.  This work relies on data products
from observations made with ESO Telescopes at the La Silla Paranal
Observatory under ESO programme ID 179.A-2005 and on data products
produced by TERAPIX and the Cambridge Astronomy Survey Unit on behalf
of the UltraVISTA consortium.  Some of the data utilized in this paper
were gathered with the 6.5 meter Magellan Telescopes located at Las
Campanas Observatory, Chile.  This paper is also based [in part] on
data collected at the Subaru Telescope and retrieved from the HSC data
archive system, which is operated by Subaru Telescope and Astronomy
Data Center at National Astronomical Observatory of Japan.  Some of
the data presented herein were obtained at the W. M. Keck Observatory,
which is operated as a scientific partnership among the California
Institute of Technology, the University of California and the National
Aeronautics and Space Administration. The Observatory was made
possible by the generous financial support of the W. M. Keck
Foundation.  The authors wish to recognize and acknowledge the very
significant cultural role and reverence that the summit of Maunakea
has always had within the indigenous Hawaiian community.  We are most
fortunate to have the opportunity to conduct observations from this
mountain.  We acknowledge the support of NASA grants HST-AR-13252,
HST-GO-13872, HST-GO-13792, and NWO grants 600.065.140.11N211 (vrij
competitie) and TOP grant TOP1.16.057.

Facilities: \facility{HST (ACS, WFC3), Spitzer (IRAC), CFHT, VISTA, Magellan
  (FourStar), Subaru (HSC), Keck:I (MOSFIRE/U142)}

{} 

\appendix

\begin{figure*}
\epsscale{0.9}
\plotone{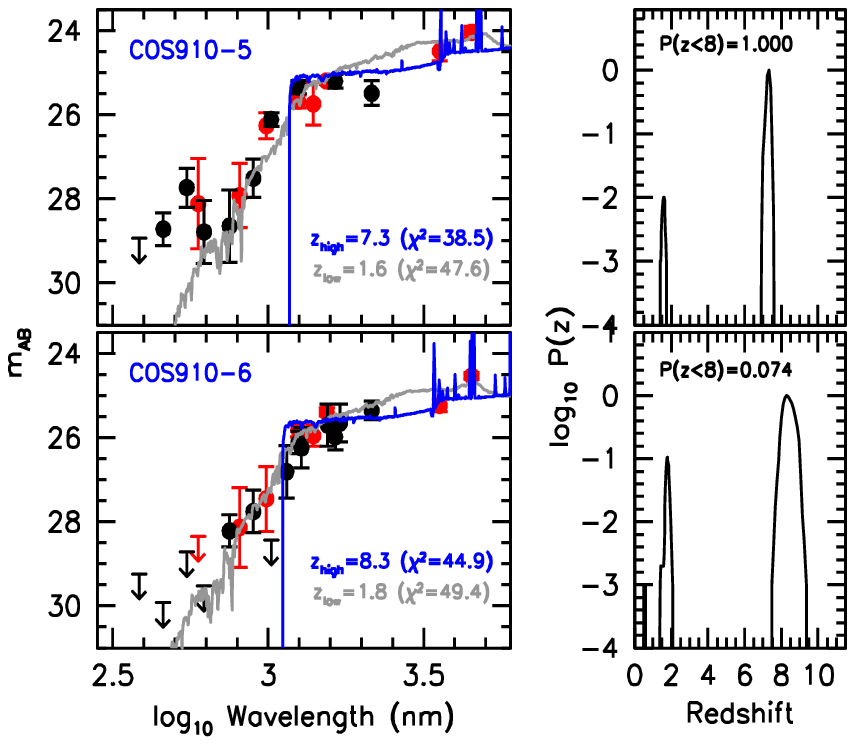}
\caption{(\textit{left panels}) Measured spectral energy distributions
  for the two candidate $z\sim9$ galaxies that we identified over the
  CANDELS fields which had measured $J_{125}-H_{160}$ colors between
  0.4 and 0.5 mag and which showed a $>$50\% probability for having a
  redshift $z>8$ using the data that were available in 2015.  The
  black solid circles show ground-based flux constraints (with
  $1\sigma$ error bars plotted), while the red points show the flux
  constraints from HST and Spitzer/IRAC.  Targeted observations of the
  two candidates have been obtained with HST in the $Y_{098}$ band
  based on a mid-cycle program (1 orbit each).  New observations are
  also available as a result of the Hyper-Suprime-Cam Ultra-deep
  observations over the COSMOS field and deeper near-IR observations
  from UltraVISTA (DR3).  The blue and gray lines show the best-fit
  $z>6$ and $z<6$ model fits, respectively, to our measured
  photometric constraints for the candidates.  (\textit{right panels})
  The redshift likelihood distributions we derive based on our
  photometric constraints for the two $z\sim9$ candidates.  While the
  first candidate COS910-5 shows roughly an equal likelihood of being
  at $z\sim1.6$ or $z\sim7.3$, the redshift of the second candidate
  COS910-6 appears to be robustly $z\sim8.3$.\label{fig:sed}}
\end{figure*}

\section{A.  Nature of Two Candidate $z\sim9$ Galaxies Targeted by HST Follow-up Program 14459}

Here we present constraints on the spectral energy distributions and
redshift likelihood distribution we derived for the two candidate
$z\sim9$ galaxies we targeted with an HST follow-up program in cycle
23 (GO 14459: Bouwens 2015).  The sources were selected for follow-up
based on their exceptional brightness ($H_{160,AB}<25.5$) and $>$50\%
probability of lying at $z>8$ (relying on our earlier photometry).
Constraints on the redshifts of the targeted candidates were derived
based on the same HST+Spitzer/IRAC+ground-based observations as
utilized by Bouwens et al.\ (2016).  We designate them COS910-5 and
COS910-6, and they have coordinates of 10:00:31.39, 02:26:39.8 and
10:00:20.12, 02:14:13.0, respectively.  These candidates had
$H_{160,AB}$ magnitudes of 25.1 mag and 25.3 mag, respectively.  The
$J_{125}-H_{160}$ colors of both sources were bluer than the $>$0.5
mag limit we had earlier used in selecting sources for follow-up
(Bouwens et al.\ 2016).

\begin{deluxetable*}{ccccccc}
\tablewidth{0pt}
\tablecolumns{7}
\tabletypesize{\footnotesize}
\tablecaption{A Small Sample of Probable\tablenotemark{a} $z\sim8.0$-8.4 Galaxies Identified over the CANDELS UDS, COSMOS, and EGS programs\tablenotemark{b}\label{tab:catalog_z80}}
\tablehead{
\colhead{ID} & \colhead{R.A.} & \colhead{Dec} & \colhead{$H_{160,AB}$} & \colhead{$z_{phot}$} & \colhead{P($z>8$)} & \colhead{P($z>7$)}}
\startdata
COS910-6 & 10:00:20.12 & 2:14:13.0 & 25.3 $\pm$ 0.1 & 8.3 & 0.87\tablenotemark{c} & 0.98\\
EGS910-11 & 14:19:59.71 & 52:51:19.5 & 26.4 $\pm$ 0.2 & 8.4 & 0.55 & 0.93\\
EGS910-12 & 14:20:19.08 & 53:03:14.3 & 25.9 $\pm$ 0.1 & 8.1 & 0.50 & 0.90
\enddata
\tablenotetext{a}{$P(z>8)>0.5$}
\tablenotetext{b}{See Appendix A and B}
\tablenotetext{c}{See Figure~\ref{fig:sed} and Appendix A.}
\end{deluxetable*}

\begin{deluxetable*}{cccccccccc}
\tablewidth{0pt}
\tablecolumns{10}
\tabletypesize{\footnotesize}
\tablecaption{Possible\tablenotemark{a} $z\sim9$-10 Galaxies Identified over the CANDELS UDS, COSMOS, and EGS programs\tablenotemark{b}\label{tab:catalog_lq}}
\tablehead{
 & & &  &  \multicolumn{3}{c}{HST+IRAC+Ground\tablenotemark{c}} & \multicolumn{3}{c}{HST+IRAC\,[3.6]/[4.5]} \\
\colhead{ID} & \colhead{R.A.} & \colhead{Dec} & \colhead{$H_{160,AB}$} & \colhead{$z_{phot}$} & \colhead{P($z>8$)} & \colhead{P($z>7$)} & \colhead{$z_{phot}$}& \colhead{P($z>8$)} & \colhead{P($z>7$)}}
\startdata
\multicolumn{10}{c}{From Appendix C and Table 7 of Bouwens et al.\ (2016)}\\
UDS910-2 & 02:17:13.08 & $-$05:15:55.4 & 26.6$\pm$0.2 & 10.2 & 0.68 & 0.91 & --- & --- & ---\\ 
UDS910-3 & 02:17:52.38 & $-$05:15:06.3 & 26.9$\pm$0.2 & 9.4 & 0.28 & 0.60 & --- & --- & ---\\ 
UDS910-4 & 02:17:14.61 & $-$05:15:15.7 & 26.6$\pm$0.2 & 9.1 & 0.50 & 0.70 & --- & --- & ---\\
\\
\multicolumn{10}{c}{Candidates where P($z>8$) $<$ 0.5}\\
COS910-7 & 10:00:26.15 & 02:32:49.8 & 25.6$\pm$0.1 & 7.4 & 0.31 & 1.00 & --- & --- & ---\\
UDS910-5 & 02:18:20.45 & $-$05:14:07.9 & 25.9$\pm$0.2 & 8.8 & 0.46 & 0.70 & --- & --- & ---\\
EGS910-13 & 14:20:45.23 & 53:02:01.4 & 25.9$\pm$0.1 & 8.5 & 0.40 & 0.95 & --- & --- & ---\\
EGS910-14 & 14:20:05.09 & 52:58:02.6 & 25.7$\pm$0.1 & 8.3 & 0.38 & 0.77 & --- & --- & ---\\
EGS910-15 & 14:19:52.22 & 52:55:58.8 & 26.1$\pm$0.1 & 8.3 & 0.38 & 0.83 & --- & --- & ---\\
EGS910-16\tablenotemark{d} & 14:20:47.81 & 53:02:11.8 & 25.6$\pm$0.1 & 8.6 & 0.15 & 0.23 & --- & --- & ---\\
\\
\multicolumn{10}{c}{Candidates Found Downweighting Constraints from the Ground-Based Photometry\tablenotemark{$\dagger$} -- Some Foreground Confusion\tablenotemark{e}}\\
COS910-8\tablenotemark{*} & 10:00:34.99 & 02:14:01.1 & 24.5$\pm$0.1 & 2.1 & 0.00\tablenotemark{*} & 0.00\tablenotemark{*} & 9.4 & 0.90 & 0.90 \\
COS910-9\tablenotemark{**} & 10:00:43.84 & 02:13:50.4 & 25.6$\pm$0.1 & 1.8 & 0.03 & 0.03 & 9.2 & 0.63 & 0.66\\
COS910-10\tablenotemark{f} & 10:00:24.22 & 02:17:56.5 & 24.4$\pm$0.1 & ---\tablenotemark{f} & --- & --- & 9.2 & 0.55 & 0.55\\
UDS910-6\tablenotemark{***} & 02:18:00.63 & $-$05:14:43.2 & 25.7$\pm$0.1 & 3.7 & 0.00 & 0.00 & 8.9 & 0.76 & 0.93\\
UDS910-7\tablenotemark{$\ddagger$} & 02:17:13.13 & $-$05:13:01.6 & 24.8$\pm$0.1 & 0.6 & 0.00 & 0.00 & 9.2 & 0.61 & 0.62\\
UDS910-8 & 02:17:25.62 & $-$05:09:38.8 & 26.1$\pm$0.3 & 9.5 & 0.31 & 0.47 & 9.1 & 0.56 & 0.77\\
UDS910-9 & 02:17:39.67 & $-$05:14:10.8 & 25.6$\pm$0.1 & 2.4 & 0.38 & 0.38 & 9.3 & 0.43 & 0.46\\
EGS910-17 & 14:20:24.88 & 53:02:35.0 & 26.5$\pm$0.1 & 9.2 & 0.55 & 0.73 & 9.2 & 0.60 & 0.78\\
EGS910-18 & 14:20:38.81 & 53:03:58.1 & 26.3$\pm$0.2 & 1.6 & 0.21 & 0.27 & 8.6 & 0.39 & 0.63\\
\\
\multicolumn{10}{c}{Candidates Found Downweighting Constraints from the Ground-Based Photometry\tablenotemark{$\dagger$} -- Minimal Foreground Confusion\tablenotemark{g}}\\
COS910-11\tablenotemark{****} & 10:00:14.78 & 02:18:09.6 & 26.2$\pm$0.1 & 1.6 & 0.02 & 0.15 & 9.0 & 0.40 & 0.45\\
COS910-12 & 10:00:24.18 & 02:26:57.2 & 25.6$\pm$0.2 & 8.6 & 0.52 & 0.63 & 8.6 & 0.43 & 0.65\\
UDS910-10\tablenotemark{**} & 02:17:38.84 & $-$05:15:29.6 & 26.1$\pm$0.1 & 1.4 & 0.00 & 0.26 & 9.4 & 0.57 & 0.60\\
UDS910-11 & 02:17:07.36 & $-$05:11:49.0 & 26.3$\pm$0.1 & 9.0 & 0.50 & 0.54 & 8.8 & 0.34 & 0.45\\
UDS910-12\tablenotemark{***} & 02:17:01.03 & $-$05:11:00.4 & 24.2$\pm$0.1 & 2.3 & 0.01 & 0.01 & 9.2 & 0.61 & 0.61\\
UDS910-13\tablenotemark{*****} & 02:17:01.36 & $-$05:09:59.6 & 25.3$\pm$0.1 & 2.3 & 0.00\tablenotemark{*****} & 0.00\tablenotemark{*****} & 9.7 & 0.68 & 0.68 \\
UDS910-14 & 02:17:49.19 & $-$05:15:16.7 & 26.9$\pm$0.2 & 9.5 & 0.68 & 0.73 & 9.5 & 0.27 & 0.29 \\
\enddata
\tablenotetext{a}{$P(z>8)>0.2$}
\tablenotetext{b}{See Appendix B}
\tablenotetext{c}{In computing the photometric solution for the 12 lowest sources in the table, we downweighted the optical photometric constraints somewhat, by multiplying their uncertainties by a factor of 2.  This was done to allow for imperfect subtraction of the optical flux from sources neighboring the indicated sources and also not to overly select against sources that show occasional $\sim$2$\sigma$ detections in the optical.}
\tablenotetext{d}{Source has a very red $H_{160}-[3.6]$ color of 1.9 mag.  Given what has been found for other sources (Figure~\ref{fig:h36}), it seems very likely that this source is at $z<4$.}
\tablenotetext{e}{Sources where subtraction of neighboring sources impacts the ground-based optical flux measurements by more than $1\sigma$ in some bands.}
\tablenotetext{f}{Ground-based photometry is substantially impacted by a bright nearby neighbor, with subtracted flux from our photometric apertures (1.2$''$-diameter) 15-20$\times$ greater than the $1\sigma$ noise in many bands.  As such, deriving accurate photometry for this source is very challenging.  In any case, this source seems consistent with being a $z<4$ dusty source given that its $H-[3.6]$ color is $\sim$1.95 mag (see Figure~\ref{fig:h36}).}
\tablenotetext{g}{Sources where subtraction of neighboring sources impacts the ground-based optical flux measurements by less than $1\sigma$ in all bands.}
\tablenotetext{$\dagger$}{Source selected if formal uncertainties on the ground-based flux measurements is increased by a factor of 10.  Because of the proximity of some HST-detected sources to nearby neighbors, obtaining accurate flux measurements for some sources is challenging.  This can result in some bona-fide $z\sim9$-10 galaxies being missed, when utilizing the ground-based photometry to determine the likely redshift of a source.}
\tablenotetext{$\ddagger$}{This source is detected at 1.9$\sigma$ and $5\sigma$ in the $I_{814}$ and $Y$ bands in our photometry.}
\tablenotetext{*}{This source is detected at $\sim$$3\sigma$ in our ground-based FourStar $J2$ photometry of ZFOURGE (Straatman et al.\ 2016: see Figure~\ref{fig:cos9108-stamps}).  As a result of this and the overall shape of the SED as defined by the existing observations in the $H$, $K_s$, [3.6], [4.5], [5.8], and [8.0] bands, this source seems very unlikely to be at $z>9$.}
\tablenotetext{**}{This source is detected at $3\sigma$ in our $Y$-band ground-based photometry}
\tablenotetext{***}{This source is detected at $2\sigma$ in our $Y$-band ground-based photometry}
\tablenotetext{****}{This source shows a $>$$3\sigma$ detection at $1.1\mu$m combining the $Y$, $J1$ and $J2$ band imaging data.}
\tablenotetext{*****}{This source is detected at $\sim$$4\sigma$ in the ground-based FourStar $J2$-band data from ZFOURGE.  Clearly, this source is not likely to be at $z>9$.}
\end{deluxetable*}

\begin{figure*}
\epsscale{1.17}
\plotone{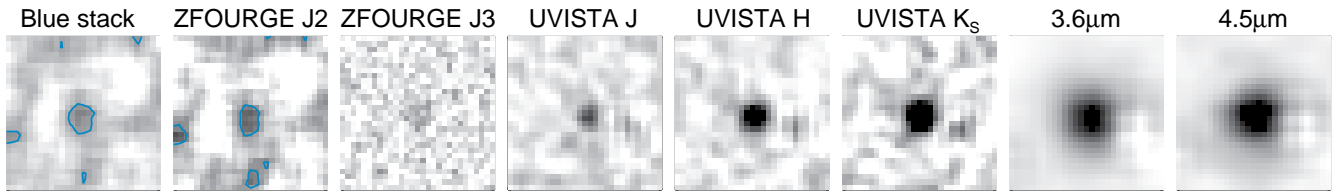}
\caption{Illustration of the neighbor-subtracted postage stamp images
  ($5'' \times 5''$) of one source COS910-8 (10:00:34.99, 02:14:01.1)
  in the ZFOURGE $J2$ and $J3$ bands (2nd + 3rd leftmost panels),
  VISTA $J$, $H$, and $K_s$ bands (middle panels), Spitzer/IRAC
  3.6$\mu$m and 4.5$\mu$m bands (rightmost panels), and in a stack of
  the imaging data blueward of 1.25$\mu$m (leftmost panels).  The blue
  stack weights the $<$1.25$\mu$m ground-based images assuming
  COS910-8 is a $z\sim2$ galaxy.  The blue contours in the leftmost
  two panels indicate regions detected at $>$2$\sigma$ significance.
  COS910-8 seemed likely to be at $z>9$ based on the HST+IRAC
  $3.6\mu$m+4.5$\mu$m photometry, but for which the ground-based
  photometry indicates is more likely at $z\sim2$ (see
  Figure~\ref{fig:cos9108}).  The presented stamps should make it
  clear that the neighboring source (separated by 1.5$''$ [mostly to
    the west, i.e., on the right-hand side] and with a $H$-band flux
  of 22.6 mag) is well subtracted.  The apparent detection of this
  source both in the FourStar J2 band from ZFOURGE and also in the
  other bluer imaging data strongly suggests that COS910-8 is not at
  $z>9$.\label{fig:cos9108-stamps}}
\end{figure*}

\begin{deluxetable}{cc}
\tablewidth{200 pt}
\tablecolumns{2}
\tabletypesize{\footnotesize}
\tablecaption{Photometry of COS910-8\tablenotemark{a}\label{tab:cos9108}}
\tablehead{\colhead{Band} & \colhead{Measured Flux (nJy)}}
\startdata
CFHTLS $u^*$  & $    16 \pm    8 $  \\
SSC $B$  & $   -11 \pm    6 $  \\
HSC $g$  & $    12 \pm    9 $  \\
CFHTLS $g$  & $   -10 \pm    7 $  \\
SSC $V$  & $     2 \pm   12 $  \\
HST $V_{606}$ & $    -7 \pm   16 $  \\
HSC $r$  & $     7 \pm    8 $  \\
CFHTLS $r$  & $    -8 \pm    9 $  \\
SSC $r^+$  & $    -1 \pm   12 $  \\
SSC  $i^+$  & $     3 \pm   16 $  \\
CFHTLS $y$  & $    19 \pm   13 $  \\
CFHTLS $i$ & $     7 \pm   10 $  \\
HSC $i$  & $    24 \pm   12 $  \\
HST $I_{814}$ & $     1 \pm   16 $  \\
CFHTLS $z$  & $   -17 \pm   25 $  \\
HSC $z$  & $    44 \pm   18 $  \\
SSC  $z^+$  & $    37 \pm   50 $  \\
HSC $y$  & $    27 \pm   43 $  \\
UVISTA $Y$  & $    35 \pm   29 $  \\
ZFOURGE $J1$ & $    12 \pm   43 $  \\
ZFOURGE $J2$ & $   152 \pm   46 $ \tablenotemark{$\dagger$} \\
HST $J_{125}$ & $   242 \pm   24 $  \\
UVISTA $J$   & $   222 \pm   33 $  \\
ZFOURGE $J3$ & $   408 \pm   63 $  \\
HST $JH_{140}$ & $   520 \pm   32$\tablenotemark{$\ddagger$} \\
HST $H_{160}$ & $   591 \pm   20 $  \\
ZFOURGE $H_s$ & $   587 \pm  119 $  \\
UVISTA $H$  & $   625 \pm   42 $  \\
ZFOURGE $H_l$ & $   601 \pm  115 $  \\
ZFOURGE $K_\mathrm{s}$ & $  1101 \pm   98 $  \\
UVISTA $K_\mathrm{s}$  & $  1063 \pm   55 $  \\
IRAC $3.6\mu$m  & $  1645 \pm   78 $  \\
IRAC $4.5\mu$m  & $  1846 \pm   61 $  \\
IRAC $5.8\mu$m  & $ -3971 \pm 1704 $\tablenotemark{*}  \\
IRAC $8.0\mu$m  & $ -2612 \pm 2363 $\tablenotemark{*} 
\enddata
\tablenotetext{a}{Figure~\ref{fig:cos9108} shows a fit of the SED of this source to $z\sim2.1$ galaxy and a $z\sim9.4$ galaxy, while Figure~\ref{fig:cos9108-stamps} shows postage stamp images of this source.}
\tablenotetext{$\dagger$}{No signal is expected in the FourStar $J2$ band from ZFOURGE (which runs from 1.07$\mu$m to 1.22$\mu$m) if COS910-8 is a $z>9$ source.  Nonetheless, COS910-8 shows a 3.3$\sigma$ detection in the $J2$ band in our photometry (see second leftmost panel in Figure~\ref{fig:cos9108-stamps}).}
\tablenotetext{$\ddagger$}{The $JH_{140}$-band flux we estimate correcting $JH_{140}$ flux measurements made in 0.4$''$-diameter apertures to total is 469$\pm$29 nJy.} 
\tablenotetext{*}{A $2\sigma$ detection of COS910-8 would be expected at both 5.8$\mu$m and 8.0$\mu$m, if it was a $z\sim9.4$ galaxy.  By contrast, if the source is instead at $z\sim2$, no detection is expected in either band due to a turn-over in the SED at 1.6$\mu$m rest-frame.  COS910-8 is not detected in either channel.}
\end{deluxetable}

As both of the targetted $z\sim9$ candidates were found over the
CANDELS COSMOS field, we made use of the same imaging data sets and
photometry as we used in selecting them.  In addition to those
photometric constraints, we also made use of the $zy$ imaging
observations associated with Hyper Suprime-Cam (HSC) Ultra-Deep survey
(DR1: Aihara et al.\ 2018a,b).  The photometric redshift software EAZY
(Brammer et al.\ 2008) is applied to the full set of flux measurements
we have for the candidates to quantify the redshift likelihood
distribution.  The best-fit high ($z>6$) and low-redshift ($z<6$) SEDs
for the sources are shown in Figure~\ref{fig:sed} with the blue and
gray lines, respectively.  The best-fit redshifts we derive for
COS910-5 and COS910-6 are 7.3 and 8.3, respectively, with $P(z<8)=1$
and $P(z<8)=0.05$.  Our first candidate COS910-5 is plausibly a lower
redshift ($z<2$) galaxy based on its photometry, while the photometry
of COS910-6 securely places it at $z>7$.

\begin{figure}
\epsscale{0.7}
\plotone{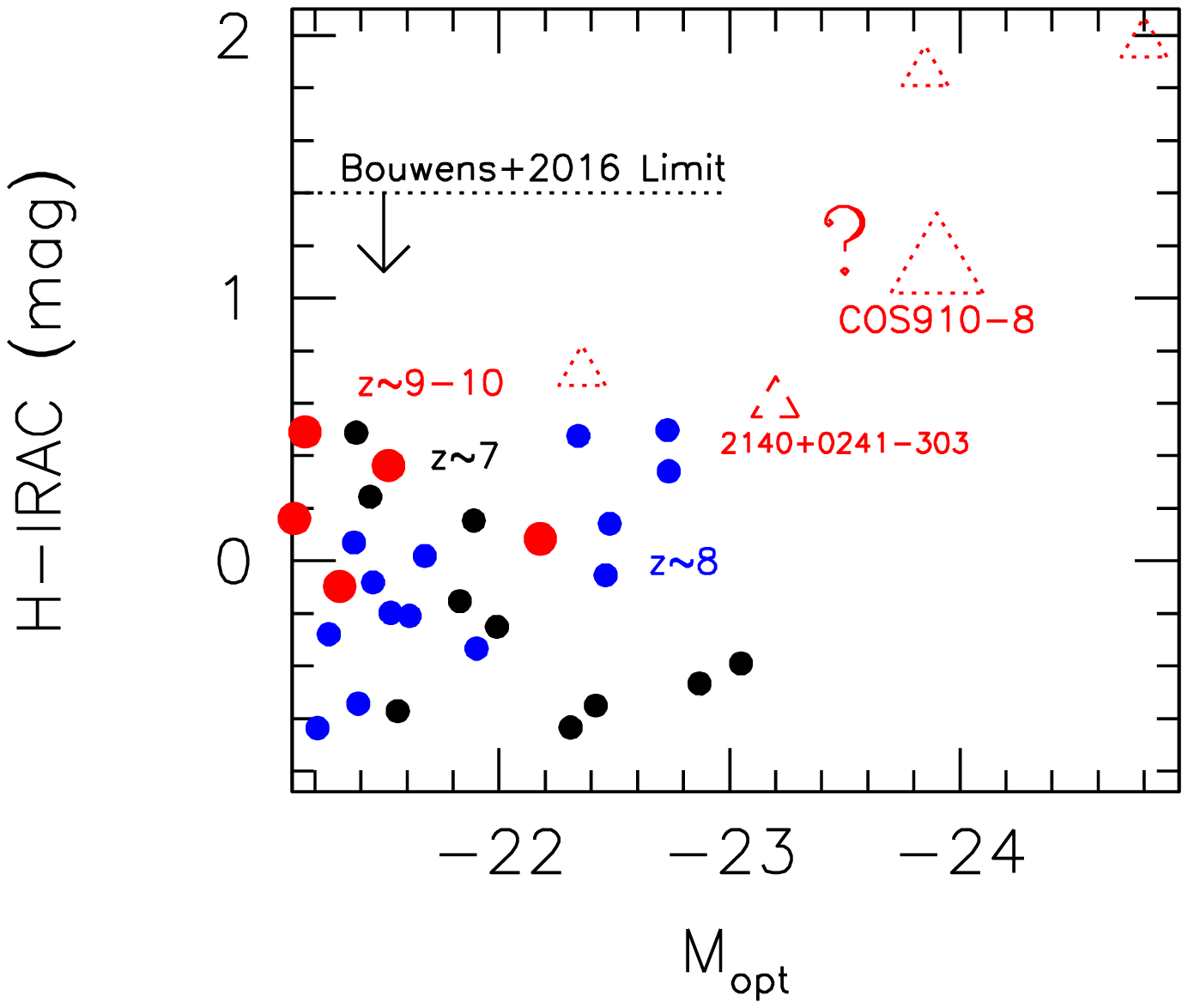}
\caption{$H - IRAC$ colors of candidate $z\sim7$-10 Galaxies
  vs. rest-optical luminosities $M_{opt}$.  The brightest $z\sim7$ and
  $z\sim8$ sources in the $>$1.5 deg$^2$ selections of Bowler et
  al.\ (2017) and Stefanon et al.\ (2019) are indicated by the solid
  black and blue circles, respectively.  $z\sim9$-10 sources
  (\textit{solid red circles}) are from Oesch et al.\ (2014) and
  Bouwens et al.\ (2016).  The dashed open triangle is a $z\sim10$
  candidate from Morishita et al.\ (2018).  The open dotted red
  triangles show the colors and nominal luminosities of candidate
  $z\sim9$-10 galaxies identified using the HST+Spitzer/IRAC
  $3.6\mu$m+4.5$\mu$m data alone and which nominally have fairly
  extreme properties relative to $z\sim7$-8 samples.  Given that we
  would tend to expect very wide-area searches such as performed by
  Bowler et al.\ (2017) and Stefanon et al.\ (2019) to yield sources
  with much more extreme properties than identified over surveys like
  CANDELS (where a $6\times$ smaller area is available) and given the
  expectations that galaxies at $z\sim7$-8 are brighter and redder
  than at $z\sim9$-10, it is odd to find candidates with even more
  extreme properties over CANDELS at $z>9$.  We might thus expect most
  of the extreme $z\gtrsim9$ candidates over CANDELS to actually be at
  $z<4$; in fact, all of the extreme candidates identified in Appendix
  B (Table~\ref{tab:catalog_lq}) appear to revert to $z<4$ solutions
  after incorporating constraints from existing ground-based
  observations.  The limiting $H_{160}-[3.6]$ color for the Bouwens et
  al.\ (2016) selection is indicated by the horizontal dotted line.
  It should be apparent that this limit easily allows for the
  selection of every bright $z\sim7$-9 galaxy identified over square
  degree fields by Bowler et al.\ (2017) and Stefanon et al.\ (2019)
  and also the $z\sim9$-10 candidates from Oesch et
  al.\ (2014).\label{fig:h36}}
\end{figure}

\section{B.  Other Potential $z\sim9$-10 Candidates}

In addition to the high-likelihood candidate $z\sim9$-10 galaxies
presented in the body of this manuscript, we also identified other
sources which also plausibly correspond to $z\sim9$-10 galaxies.
However, because the integrated probability of these sources lying in
excess of 8.0 was below 50\% or because their maximum likelihood
redshift was below 8.4, we excluded them from our primary sample.

In an effort to be comprehensive, we include in our compilation even
bright sources which showed at least a 20\% probability of lying at
$z>8$.  We furthermore relaxed the $H-[3.6]<1.4$ mag criterion used
for our main selection to include sources redder than 1.4 mag, but as
we discuss in Appendix B.2 and illustrate in Figure~\ref{fig:h36},
this does not appear to add any probable $z\sim9$-10 candidates.

Table~\ref{tab:catalog_z80} provides a list of sources where the best
estimate redshift is between 8.0 and 8.4, with a $>$50\% probability
of lying at $z>8$.  Table~\ref{tab:catalog_lq} features a list of
sources where there is a $>$20\% probability of lying at $z>8$.

For completeness, we have also included in this table sources which
were previously listed in Table 7 and Appendix C of Bouwens et
al.\ (2016).

\subsection{B.1  Possible $z\sim9$-10 Candidates Identified By Down-Weighting the Ground-Based Data}

A fraction of the sources ($\sim$30\%) found in the HST data are very
close by other sources in the ground-based imaging observations.
While our photometric software \textsc{mophongo} copes with source
overlap (subtracting flux from nearby neighbors before doing
photometry), subtractions are not perfect in all cases and this can
cause some fraction of bona-fide high-redshift sources to be missed
due e.g. to imperfectly subtracted optical flux from neighboring
sources.  Based on the selection volume simulations run in \S4.1, we
estimate this incompleteness to be $\sim$30\%.

Given these incompleteness levels in HST+Spitzer+ground-based
$z\sim9$-10 selections, we repeated our $z\sim9$-10 galaxy searches
using the same procedure as in the main text but increasing the
uncertainties on the ground-based data by a factor of 10 (and hence
significantly downweighting those data).  Those candidates are
included in Table~\ref{tab:catalog_lq} along with their best-fit
redshifts and integrated probabilities of lying at $z>7$ and $z>8$.
For comparison, we include in the same table the maximum likelihood
redshifts and the integrated probabilities of sources lying at $z>7$
and $z>8$ when including all the data in the middle columns.  This is
to show the impact of incorporating the photometric constraints
available from the ground-based $Y$, $J1$, $J2$, $J3$, $J$, $H_s$,
$H$, $H_l$, and $K$ band observations.  The optical photometric
constraints are also included in the results presented in the middle
column, but the photometric uncertainties there are multiplied by a
factor of 2 to allow for the possibility of imperfectly subtracted
neighbors.  Sources with best-fit redshifts $z>8$ and where the
integrated probability at $z>8$ is $>$20\% are also included in
Table~\ref{tab:catalog_lq}.

\subsection{B.2  Case of $z\sim9$ Candidate COS910-8}

There are a few sources in Table~\ref{tab:catalog_lq} which seem like
reliable $z\sim9$-10 candidate galaxies when using only the
HST+Spitzer/IRAC 3.6$\mu$m+4.5$\mu$m observations, but which clearly
prefer a $z<4$ solution when including the constraints from the
ground-based photometry.  In Figure~\ref{fig:cos9108} from the main
text, we showed one example of this type of source which prefer a
$z>9$ solution using the HST+3.6$\mu$m+4.5$\mu$m observations alone.
Because of the brightness of that source and to ensure that we were
not deriving a low-redshift solution due to imperfect subtraction of a
bright nearby source, we took unusual care in subtracting a bright
neighboring source before performing our flux measurements
(1.2$''$-diameter apertures).  Postage stamp images of this source,
after subtraction of a neighboring source, are shown in
Figure~\ref{fig:cos9108-stamps}.  The leftmost image in this panel
show a weighted stack of all the data from the $U$-band to the
FourStar J2 band, weighted according to expected contribution in each
band if the source were a $z\sim2$ galaxy.

Particularly concerning for COS910-8 being a $z>9$ source is the
apparent detection of this source at $3\sigma$ in the FourStar J2 band
at the position of the source, tentative detections of the source in
the $i$, $z$, and $Y$ bands (14$\pm$6 nJy, 24$\pm$14 nJy, and
27$\pm$21 nJy, respectively), and the overall SED shape as defined by
the available $H$, $K_s$, [3.6], [4.5], [5.8], and [8.0]-band
photometry.  In particular, the measured $K_s$-band flux from both
UltraVISTA and FourStar/ZFOURGE lie in excess of expectations if the
source is at $z>9$, while our 5.8$\mu$m and 8.0$\mu$m flux constraints
lie below expectations if the source is at $z>9$.  Including the
photometric constraints from all HST, Spitzer, and ground-based data
(Table~\ref{tab:cos9108}), we estimate that the source has a
2$\times$10$^{-6}$ probability of lying at $z>9$.  If slightly larger
photometric uncertainties are assumed in all bands (as needed to
derive a reduced $\chi^2$ of 1), the integrated likelihood that the
source is at $z>9$ is 2$\times$10$^{-4}$.  

Consistent with our interpretation that COS910-8 is much more likely a
$z\sim2$ galaxy than a $z\sim9.4$ galaxy, we find no significant line
emission ($<$$4.2\times10^{-18} \textrm{ergs/s/cm}^2$) in the Keck
observations we obtained of COS910-8 in the $H$ band at 1.6$\mu$m
which probe \CIV\ $\lambda$1548,1550, \HeII, and \OIII\ $\lambda$1663.
Line emission should be detected at $>$5$\sigma$ in our observations
if the EW of any of these lines was in excess of 10\AA, as has already
reported in a $z=6.11$ and $z=7.045$ galaxy by Schmidt et al.\ (2017)
and Stark et al.\ (2015).

Besides COS910-8, there are 7 other candidates in
Table~\ref{tab:catalog_lq}, which strongly prefer a low-redshift
solution after incorporating the constraints from ground-based
observations, mostly as a result of a detection in a band just
blueward of $1.2\mu$m, i.e., Y, J1, or J2.  Given the impact of the
optical data on the computed $P(z>8)$'s, clearly it is essential to
carefully examine the ground-based optical and near-IR observations
when selecting the best $z>8$ galaxies for further follow-up
observations.

Even without a detailed consideration of a vast array of ground-based
observations to determine the nature of COS910-8 and a number of
similar sources within CANDELS, it is clear that we might have
expected most of these sources to correspond to $z<4$ galaxies given
the very high implied luminosities and red $H$$-$IRAC colors (treating
them as $z>9$ candidates).  To illustrate how extreme e.g. COS910-8
would be as a $z>9$ candidate, we present its optical luminosity and
$H$$-$[3.6] color (assuming a $z\sim9.4$ redshift: \textit{red open
  dotted triangle}) relative to the $z\sim7$ and $z\sim8$-9 samples
identified by Bowler et al.\ (2017) and Stefanon et al.\ (2019) over
$>$1.5 deg$^2$ in Figure~\ref{fig:h36}.  The optical luminosities and
$H$-IRAC colors from the Bowler et al.\ (2017) and Stefanon et
al.\ (2019) selections only rely on IRAC bands not impacted by the
[OIII] or H$\alpha$ nebular emission lines and hence only rely on
sources at $z\geq 6.65$.  As such, the luminosities are computed from
the IRAC 4.5$\mu$m band for photometric candidates identified in the
redshift range $z=6.65$-6.95 and from the IRAC 3.6$\mu$m band for
sources at redshifts $z>7$.

Given the $6\times$ larger area of the Bowler et al.\ (2017) and
Stefanon et al.\ (2019) searches relative to the current CANDELS
probes and given that bright, red sources are presumably more common
at $z\sim7$-8 than at $z\sim9$-10, it would seem odd to find more
extreme sources in a $z\sim9$-10 search over CANDELS than in the
aforementioned lower-redshift, wider-area probes, and it therefore
seems probable that most of the extreme candidates over CANDELS are
actually at $z<4$.

Also shown in Figure~\ref{fig:h36} are the luminosities and colors of
the $z\sim9$-10 galaxies in the Oesch et al.\ (2014), Bouwens et
al.\ (2016), and Morishita et al.\ (2018) selections and the implied
luminosities and colors for the other $z>8$ candidates from
Table~\ref{tab:catalog_lq} (\textit{red open dotted triangles}) that our
ground-based photometry largely rule out.  The red $H-$IRAC color
selection limit used by Bouwens et al.\ (2016) for preselecting their
$z\sim9$-10 samples is shown in Figure~\ref{fig:h36}, and it is clear
that such a selection would easily include all extreme $z\sim7$-10
sources identified by Bowler et al.\ (2017), Stefanon et al.\ (2019),
and Oesch et al.\ (2014).

The purpose of the entire discussion here is to illustrate the
challenges present in selecting high-quality $z>8$ samples based on
current data sets and how important accurate photometry is.

\subsection{B.3  Possible Impact of these $z\sim9$-10 Candidates on Our LF Estimates?}

If we add up the fractional probability that the 28 sources presented
in Table~\ref{tab:catalog_lq} and \ref{tab:catalog_z80} are at $z>8$,
i.e., $P(z>8)$, at face value this suggests that 8.67 of the presented
candidates are at $z>8$.  As this constitutes of a non-negligible
fraction of the 19 candidates presented in Table
~\ref{tab:catalog_conf}, one might wonder whether the presented
candidates in Appendix A and B would have an impact on the LF if they
were indeed at $z\sim9$-10.

As we argue below, the answer is that we would not expect the
presented candidates to impact our derived LFs significantly, and the
reason is that we expect to miss 30\% of the bona-fide $z\sim9$-10
candidates using the selection criteria utilized in this paper.  Such
a completeness fraction is built into the selection volume estimates
we estimate for the CANDELS UDS, COSMOS, and EGS fields.  The volumes
we estimate for those fields are 30\% lower than for fields like
CANDELS GOODS-North where deeper ACS/optical and $Y_{105}$-band
coverage is available.  In most cases, this incompleteness occurs due
to a blending of some $z\sim9$-10 sources with neighboring sources and
some limitations in the sensitivity and wavelength coverage available
over the CANDELS WIDE fields.

Let now estimate how many bona-fide $z\sim9$ candidates we are likely
to find if we follow up all the candidates from
Tables~\ref{tab:catalog_lq} and \ref{tab:catalog_z80} with deeper
$Y_{105}$-band observations.  Including only those candidates from
Tables~\ref{tab:catalog_lq} and \ref{tab:catalog_z80} which have a
redshift $z>8.4$, the total fractional probability drops from 8.67 to
5.68.  

In addition, we should calibrate our estimated $P(z>8)$'s based on the
follow-up results from Bouwens et al.\ (2016) which confirmed only
50\% of 9 $P(z>8)>0.5$ candidates as bona-fide $z>8$ source (see our
discussion in \S3.2).  As the $P(z>8)$'s before follow-up sum to 6.5
in Bouwens et al.\ (2016) but were found to be 4.5 based on the
follow-up data, we should renormalize the computed $P(z>8)$'s down by
31\%.  Incorporating both effects, the total fractional probability is
3.91.  In other words, we would expect to find 4 more bona-fide
$z\sim9$ candidates by following up all the sources in
Tables~\ref{tab:catalog_lq} and \ref{tab:catalog_z80} with
$Y_{105}$-band observations.

There are 9 candidate $z\sim9$-10 galaxies from the CANDELS UDS,
COSMOS, and EGS fields given in Table~\ref{tab:catalog_conf} that we
use to derive our $z\sim9$-10 LF results.  After we account for the
insecure nature of 5 of them (counting each as 0.5), the total is 6.5.
If we compare the effective number of sources used for the present LF
determinations, i.e., 6.5 with the total sample including these
additional sources, i.e., 6.5+3.91 = 10.46, the fraction of sources in
our main samples is 62\%, which is not especially different from the
$\sim$70\% fraction expected from our selection volume simulations.

\section{C.  Constraints on the Nature of COS910-8 from Keck Observations}

While we were still evaluating COS910-8's viability as a possible
$z\sim9.4$ candidate on the basis of the HST+Spitzer/IRAC data
(Figure~\ref{fig:cos9108}) and the ground-based data
(Figure~\ref{fig:cos9108-stamps}), we obtained sensitive Keck MOSFIRE
(McLean et al.\ 2012) spectroscopy on this source in parallel with a
small sample of other candidate $z\sim6$-9 and $z\sim2$ galaxies in
CANDELS COSMOS.  Given the brightness of COS910-8, it may not be
especially difficult to detect line emission from the source if it
were at $z>9$.  Of the various $UV$ lines accessible to detection in
COS910-8, the \CIV\ $\lambda$1548,1550 doublet may be the easiest to
detect, if the recent results by Stark et al.\ (2015), Mainali et
al.\ (2017), and Schmidt et al.\ (2017) be any guide.  If the source
was at $z\sim9.4$, the \CIV\ $\lambda$1548,1550 doublet would fall in
the MOSFIRE $H$ band.  We carried out observations on the 23th and
24th of April 2019 as part of Keck program (U142), for a total of 4.8
hours on source integration time at $0.4''-0.8''$ seeing in the $H$
band.  The data was reduced using the public MOSFIRE
DRP\footnote{https://keck-datareductionpipelines.github.io/MosfireDRP/}
and perform telluric corrections and flux calibrations using a custom
built tool (T. Nanayakkara et al.\, in prep; Nanayakkara et
al.\ 2016).

Given our redshift estimate of $z=9.4\pm0.2$ for COS910-8 (when
reliance was made on the HST+Spitzer/IRAC information alone), we
looked especially carefully in the wavelength interval
15810\AA--16430\AA$\,\,$for possible line emission as might be
indicative of a \CIV\ 1548,1550 doublet.  Our observations reach a
typical $3\sigma$ RMS of $\sim5\times10^{-19}$
$\textrm{ergs/s/cm}^2/\textrm{\AA}$ for an empty sky region in the
slit in this redshift window. Within such limits, we find no emission
line features for COS910-8.  For a typical resolution element of 3
pixels (5\AA) on MOSFIRE, this results in an approximate $5\sigma$
detection limit for \CIV\ of $4.2\times10^{-18} \textrm{ergs/s/cm}^2$
for either line in the doublet.  If we convert this line detection
limit to the equivalent EW of \CIV\ where we would expect a $5\sigma$
detection in our data, the EW limit is 5.4\AA.  Given the likely mass
of the source if it were at $z\sim9.4$ (see Figure~\ref{fig:h36}), the
line width would seemingly be at least $\sim$200 km/s, reducing our
line sensitivity by approximately a factor of 2 relative to the case
that the line was unresolved.  If the \CIV\ doublet were as strong as
it was measured in the $z=6.11$ galaxy RXCJ2248-ID3 (Schmidt et
al.\ 2017; Mainali et al.\ 2017) where an EW of 24$\pm$4\AA$\,\,$ was
measured or A1703-zD6 (Stark et al.\ 2015) with an equivalent width of
$\sim$38\AA, \CIV\ should easily be detected (again if it were the
case that COS910-8 was indeed at $z>9$).

Assuming \CIV\ to be a single broad line, using the MOSFIRE sky
spectrum\footnote{https://www2.keck.hawaii.edu/inst/mosfire/sky\_lines.html},
we find that $\sim10\%$ of the spectral coverage falls within sky
lines with $f_{\nu}>5\times10^{-26}
\textrm{ergs/s/cm}^2\textrm{/Hz/arcsec}^2$.  Thus, there is a
$\sim10\%$ probability for \CIV\ to fall within a strong sky line
region affecting its spectroscopic identification. However, owing to
the fact that \CIV\ is a doublet and given its expected breadth (e.g.,
Laporte et al.\ 2017; Mainali et al.\ 2018), sky lines are unlikely to
strongly influence the identification of the \CIV\ doublet, if
COS910-8 was indeed at $z\sim9.4$.

We also looked briefly at constraints that can be placed on line
emission from \HeII$\,\,$and \OIII\ $\lambda$1663 assuming the source
is at $z>9$ and the constraints are similar.

\end{document}